\documentclass[conference]{IEEEtran}

\ifCLASSINFOpdf
\else
\fi

\usepackage{algorithmic}
\usepackage{graphicx}
\def\BibTeX{{\rm B\kern-.05em{\sc i\kern-.025em b}\kern-.08em
    T\kern-.1667em\lower.7ex\hbox{E}\kern-.125emX}}
    
\usepackage{xspace}
\usepackage{lipsum}
\usepackage{subfigure} 
\usepackage{diagbox}
\usepackage{color, xcolor}
\usepackage[linesnumbered,ruled]{algorithm2e}
\usepackage{booktabs}
\usepackage{multirow}
\usepackage{threeparttable}
\usepackage{array}
\usepackage{bbding}
\usepackage{threeparttable}

\usepackage{epigraph} 
\usepackage{balance}
\usepackage{microtype}
\usepackage{amsmath,amssymb,amsfonts}

\newcommand{\sys}{{\sc Raconteur}\xspace}

\usepackage{tcolorbox}
\usepackage{listings}
\usepackage{pifont}
\usepackage{makecell}
\usepackage{url}
\usepackage{soul}
\usepackage{bbm}
\definecolor{mycolor}{RGB}{241, 242, 243}
\sethlcolor{mycolor}
\newcommand{\codeword}[1]{\hl{\strut\texttt{#1}}}

\usepackage{caption}[article]
\definecolor{mycustomgreen}{RGB}{50, 220, 50}
\definecolor{mycustomred}{RGB}{193, 18, 31}
\usepackage{breakurl}
\usepackage[breaklinks]{hyperref}
 
\hypersetup{
colorlinks=true,
linkcolor=mycustomgreen,
citecolor=mycustomred,
}

\makeatletter
\def\footnoterule{\relax%
  \kern-3pt
  \hbox to \columnwidth{\vrule width 1\columnwidth height .5pt\hfill}
  \kern3pt}
\makeatother

\begin{document}
%
\title{\sys: A Knowledgeable, Insightful, and Portable LLM-Powered Shell Command Explainer}

\author{\IEEEauthorblockN{Jiangyi Deng\textsuperscript{1\ding{171}}\thanks{\ding{171}~Both authors contributed equally to the paper.}, Xinfeng Li\textsuperscript{1\ding{171}}, Yanjiao Chen\textsuperscript{1\ding{41}}, Yijie Bai\textsuperscript{1}, Haiqin Weng\textsuperscript{2}, \\Yan Liu\textsuperscript{2}, Tao Wei\textsuperscript{2}, Wenyuan Xu\textsuperscript{1}}
\IEEEauthorblockA{\textsuperscript{1}{Zhejiang University}, \textsuperscript{2}{Ant Group}\\
\{\textit{jydeng, xinfengli, chenyanjiao, baiyj, wyxu}\}@zju.edu.cn \\ \{\textit{haiqin.wenghaiqin, bencao.ly, lenx.wei}\}@antgroup.com}
}



\IEEEoverridecommandlockouts
\makeatletter\def\@IEEEpubidpullup{6.5\baselineskip}\makeatother
\IEEEpubid{\parbox{\columnwidth}{
		Network and Distributed System Security (NDSS) Symposium 2025\\
		23–28 February 2025, San Diego, CA, USA\\
		ISBN 979-8-9894372-8-3\\
		https://dx.doi.org/10.14722/ndss.2025.23798\\
		www.ndss-symposium.org
}
\hspace{\columnsep}\makebox[\columnwidth]{}}

\maketitle

\begin{abstract}

Malicious shell commands are linchpins to many cyber-attacks, but may not be easy to understand by security analysts due to complicated and often disguised code structures. Advances in large
language models (LLMs) have unlocked the possibility of generating understandable explanations for shell commands. However, existing general-purpose LLMs suffer from a lack of expert knowledge and a tendency to hallucinate in the task of shell command explanation. 
In this paper, we present \sys, a knowledgeable, expressive and portable shell command explainer powered by LLM. \sys is infused with professional knowledge to provide comprehensive explanations on shell commands, including not only what the command does (\textit{i.e.}, behavior) but also why the command does it (\textit{i.e.}, purpose). To shed light on the high-level intent of the command, we also translate the natural-language-based explanation into standard \emph{technique \& tactic} defined by MITRE ATT\&CK, the worldwide knowledge base of cybersecurity. To enable \sys to explain unseen private commands, we further develop a documentation retriever to obtain relevant information from complementary documentations to assist the explanation process. We have created a large-scale dataset for training and conducted extensive experiments to evaluate the capability of \sys in shell command explanation. The experiments verify that \sys is able to provide high-quality explanations and in-depth insight of the intent of the command. 

\end{abstract}
\section{Introduction}

Ongoing cyber-attacks are significant threats to an organization's data, applications and other valuable assets~\cite{Vielberth20Security, Kinyua21AI, Khalili19Monitoring}. Malicious shell commands often act as springboards for attackers to gain remote control of victim computers. To thwart such attempts, intrusion detection systems (IDS) usually capture and alert suspicious shell command for further investigation by human analysts. Manual auditing on alerted shell commands is a key part of security operation. 
Security analysts are expected to fully understand the commands, from their literal semantics to high-level intents. However, the complex syntax and stealthy nature of malicious commands make it difficult for junior technicians and even security experts to understand.

The recent breakthrough in large language models (LLMs) has unlocked the potential of code explanation for programmers and non-programmers. It has been found that commercial LLMs (\textit{e.g.}, ChatGPT) can generate understandable and fairly accurate explanations for commonly-used commands via user prompts~\cite{Sarsa22Automatic, macneil23experiences, MacNeil22Generating, Denny22Robosourcing, Leinonen23comparing}. Nonetheless, when presented with complicated or unseen commands, general-purpose LLMs will provide erroneous and even hallucinated explanations. This situation may be worsened for companies that use special shell commands containing private utilities or sensitive parameters.
For example, Samsung has banned the use of ChatGPT and other AI-powered chatbots after an accidental leakage of sensitive internal source code~\cite{samsung}. 

In this paper, we develop an LLM-powered shell command explainer, named \sys, which provides knowledgeable and insightful explanations on shell commands, especially malicious ones. \sys is also portable to private shell commands that are absent from its training dataset. These design goals are realized through addressing the following challenges.

\begin{itemize}
\setlength{\itemsep}{10pt}
    \item \textit{How to endow a general-purpose LLM with expert knowledge in shell command explanation?}
\end{itemize}

Existing general-purpose LLMs have unsatisfying performance on shell command explanation, especially for malicious commands. Fine-tuning a general-purpose LLM is a promising and affordable solution, but there is a lack of domain data for the task of shell command explanation. To tackle this problem, we construct a fine-tuning dataset of $\left<prompt,~response\right>$ pairs. To account for diversity in user prompts, we design both rule-based and model-based methods to generate a large variety of prompts. The $response$s are composed with professional knowledge from prestigious code libraries. After fine-tuning, \sys is able to explain the action of each step in the shell command as well as summarize the overall behavior of the shell command.

\begin{itemize}
\setlength{\itemsep}{10pt}
    \item \textit{How to provide insightful interpretations of the  intent of a shell command?}
\end{itemize}

An in-depth understanding of commands not only includes step-by-step description of the semantics but also an abstraction of the intents, \textit{i.e.}, to understand what the attacker wants to achieve (\textit{tactic}) by what means (\textit{technique}). Translating the natural-language-based description into the \emph{technique \& tactic} defined by the MITRE ATT\&CK framework helps analysts quickly seek for corresponding mitigations. For example, the semantic explanation of the command \codeword{rundll32.exe keymgr, KRShowKeyMgr} may be ``The command executes the exported function \codeword{KRShowKeyMgr} located in \codeword{keymgr.dll} using \codeword{rundll32.exe}.'' It is more helpful if analysts also understand the technique behind this command, \textit{i.e.}, to export the stored Windows credentials from the credential manager to a file, which belongs to the ``T1003 - OS Credential Dumping'' technique. The ultimate intent of this action is ``TA0006 - Credential Access'' tactic in the MITRE ATT\&CK framework. By pinpointing the technique and tactic, a wealth of prior knowledge can be harnessed to respond to this threat. Unfortunately, there is a gap between the natural-language-based explanation and the standard description given by MITRE ATT\&CK documentations. To address this gap, we establish an embedding model that maps the description of \sys and that of MITRE ATT\&CK into the same embedding space, based on which we match the command to the technique and tactic with the most similar descriptions.

\begin{itemize}
\setlength{\itemsep}{10pt}
    \item \textit{How to provide accurate information about private shell commands that are unseen in the training phase?}
\end{itemize}

A large portion of commands executed in an organization are proprietary, involving utilities, parameters, and files that are developed within the organization and are for internal use only. An LLM may provide inaccurate and fictitious information when presented with unseen commands that are not publicly available. To enable \sys to be portable to private shell commands, we design a documentation retriever to glean relevant contexts in complementary documentations (\textit{e.g.}, private documentations of the company) to assist the explanation process of \sys. In this way, \sys can analyze commands according to the provided documentations and provide faithful explanations.

We implement a fully-functional prototype of \sys and create a dataset to conduct extensive experiments to evaluate the capability of \sys. We compare \sys with three baseline models, \textit{i.e.}, GPT-3.5-Turbo, GPT-4 and ChatGLM2-6B. Among them, GPT-4 is the most powerful commercial LLM so far. Our extensive experiments demonstrate that \sys achieves superior explanation performance than the baselines in both English and Chinese languages. \sys also performs well in technique and tactic identification, which helps shed light on the high-level intents of commands.

We summarize our main contributions as follows:
\begin{itemize}
\setlength{\itemsep}{5pt}
    \item We propose \sys, an LLM-powered shell command explainer that can provide knowledgeable and insightful descriptions on shell commands, especially malicious ones, to assist security analysts in identifying potential cyber-attacks.
    \item We equip \sys with a holistic toolkit including behavior explainer, intent identifier, and documentation retriever, allowing \sys to provide comprehensive and faithful explanations on public and private shell commands. 
    \item We have conducted extensive experiments to verify that \sys is able to provide high-quality explanations and in-depth insight of the intent of the command. The dataset we curated is open-source to boost further research in LLM-aided code explanation\footnote{\url{https://raconteur-ndss.github.io/}}.
\end{itemize}

\section{Background and Motivation}

\subsection{Shell Command Explanation}\label{subsec:shell}

Shell commands allow users to interact with operating systems, \textit{e.g.}, enter commands and receive responses. Many crucial cyber-attacks, \textit{e.g.}, masquerade attacks, are enabled by malicious shell commands~\cite{Schonlau00Detecting}. Intrusion detection systems (IDS) often capture and report suspicious shell commands to security analysts for further confirmation. However, the complicated formations of shell commands usually make it difficult for junior technical personnel or even experts to truly understand its ultimate purpose. A shell command consists of \textit{utilities} (a.k.a., \textit{executables}) followed by \textit{options} and \textit{parameters}. As shown in the following benign and malicious shell commands, \codeword{bash} is a utility, \codeword{-c} is an option, and the entire subsequent string within the single quotes is a parameter.
{\small
\begin{lstlisting}[title={A Benign Shell Command}, language={}, basicstyle=\tt,
frame=shadowbox, rulesepcolor=\color{red!10!green!10!blue!10}, breaklines=true, showstringspaces=false, columns=flexible, morekeywords={ip_addr, port}]
bash -c 'exec 137<>/dev/tcp/ip_addr/port && echo -e "GET /_health/HTTP/1.1\r\nhost: ip_addr" >&137 && grep ok -s -m 1 <&137'
\end{lstlisting}
\vspace{-5pt}
\begin{lstlisting}[title=A Malicious Reverse Shell Command, language={}, basicstyle=\tt,
frame=shadowbox, rulesepcolor=\color{red!10!green!10!blue!10}, breaklines=true, showstringspaces=false, columns=flexible, morekeywords={ip_addr, port}]
bash -c '0<&137-;exec 137<>/dev/tcp/ip_addr/port;sh <&137 >&137 2>&137'
\end{lstlisting}
}
\vspace{5pt}

\emph{\underline{Technique \& tactic}}. To analyze shell commands, especially malicious ones, it is essential to understand both the technique and the tactic. The \emph{technique} is the concrete actions of the commands, \textit{i.e.}, ``how'' the commands are executed step by step. The tactic is the ultimate purpose of the commands, \textit{i.e.}, ``why'' the commands are performed. Take a shell command for running malicious codes as an example. The tactic is executing the malicious codes. In order to achieve this goal, the technique may involve abusing Unix shell scripts for execution, \textit{e.g.}, the reverse shell example above. Identifying both the tactic and the technique of a shell command helps analysts better understand the purpose and the habits of attackers. The most commonly-used tactics and techniques of malicious commands are taxonomized in MITRE ATT\&CK's\footnote{\url{https://attack.mitre.org/}}, a globally-accessible knowledge base according to real-world observations. MITRE ATT\&CK defines 14 tactics and 
about 200 techniques, which can be leveraged to interpret malicious shell commands.

\emph{\underline{Automatic command explanation}}. To assist human experts in understanding the technique and tactic of a shell command, automated explanation systems have been developed. Current breakthrough in large language models (LLMs) also gives rise to LLM-based command explanation algorithms. However, there are two main challenges facing existing systems. First, LLM-based systems suffer from intrinsic defects, especially regarding \textit{faithfulness} and \textit{factuality}. The output of an LLM-based system may not  faithfully follow user instructions (\textit{i.e.}, the answer is irrelevant) or provide true facts (\textit{i.e.}, the answer is incorrect)~\cite{Maynez20Faithfulness}. Second, the queried shell commands may be unseen to the explanation system trained with public datasets. For example, many companies design special shell utilities, \textit{e.g.,} used within the enterprise. The shell parameters, \textit{e.g.}, IP address (\codeword{ip\_addr}) and port (\codeword{port}), may be private information of a company. These special shell commands are usually absent from the training dataset of a general command explanation system, leading to poor generalization problems.  

\subsection{Generative Language Models}\label{ssec:language}
Generative language models output texts given an input \emph{prompt}, widely applied to translation, summarization, and question answering. With recent advances of large-scale pre-training, commercial LLMs such as ChatGPT exhibit remarkable performance across a broad spectrum of tasks~\cite{Brown20Language}.

A typical training pipeline of generative LLMs includes pre-training and fine-tuning~\cite{Ouyang22Training}. Due to its prohibitive requirements of training data and computational resources~\cite{Kaplan20Scaling}, pre-training is usually performed by professional AI companies like OpenAI~\cite{Ouyang22Training}. Fine-tuning is an important way of turning a general-purpose LLM into a domain-specific LLM, \textit{e.g.}, for malicious command analysis. Specifically, a pre-trained LLM can be fine-tuned on a labelled dataset of $\left<prompt,~response\right>$ pairs for the target task in the given domain. In this way, the domain-specific LLM will achieve better performance on specific tasks.  

\textit{Faithfulness} and \textit{factuality} are two serious concerns about LLMs, which means the output may diverge from the prompt (violating \textit{faithfulness}) or misalign with established world knowledge (violating \textit{factuality})~\cite{Maynez20Faithfulness}. 
Pre-trained models, \textit{e.g.}, ChatGPT, are shown to have a high degree of faithfulness but not factuality~\cite{Li23Evaluating}. To enhance factuality, retrieval-augmented generation (RAG)~\cite{Wang23CodeT5p} has been proposed, which complements the original user prompt with related facts and knowledge via web browsing~\cite{Nakano21WebGPT, Liu23WebGLM} or document retrieval~\cite{ChatGPTplugins}. The original user prompt is encoded into an embedding vector, which is used to retrieve relevant texts from available sources based on semantic similarity. The semantic similarity between two texts can be computed by measuring the distance between two embedding vectors. The user prompt and the retrieved texts are combined as a new prompt to be fed into the LLM to elicit more factual output.

\subsection{A Real-World Motivating Example}

In this part, we give an example of explaining a real-world malicious shell command by \sys. The command is presented in the above malicious reverse shell command example in \S\ref{subsec:shell}. This malicious command can enable an attacker to gain remote control, which was captured by the IDS in real-world security operation.

The aim of the malicious shell command is to establish an illegal connection to a remote computer at \codeword{ip\_addr:port} and redirect the input and output of the victim system's shell so that the attacker can access it remotely. As we can see in the example, the malicious shell command has a similar form to the benign shell command, \textit{e.g.}, they both have IP address and port and redirect the standard input and output. To make matters worse, the malicious shell command contains obfuscating file descriptors (\textit{e.g.}, \codeword{\&137}) and complicated redirections (\textit{e.g.}, \codeword{<\&137 >\&137 2>\&137}). Therefore, it is hard for analysts to identify the ulterior motive of the command.

The explanation generated by \sys for this command is shown in the following example. The explanation consists of two sections, \textit{i.e.}, behavior explanation and intent prediction. The behavior explanation section first describes the behavior of each action in the command (\textit{i.e.}, \textbf{Step by step explanation}), and then summarizes the overall behavior of the command (\textit{i.e.}, \textbf{Overall}). The intent prediction section (\textit{i.e.}, \textbf{Tactic} \& \textbf{Technique}) derives the technique and tactic of the command in a standard framework based on its behaviors.
The holistic explanation provided by \sys enables analysts to quickly comprehend both semantics and intents of the command, assisting them in making informed decisions for further countermeasures. 

\begin{tcolorbox}[title = {An Explanation on the Malicious Shell Command}, colback=white]
\small
\textbf{Step by step explanation:}
\begin{itemize}
\setlength{\itemsep}{2pt}
    \item \codeword{bash -c}: This command is invoking the Bash shell and running a command within it.
    \item \codeword{0<\&137-}: This is redirecting file descriptor 0 (stdin) to file descriptor \codeword{137} and closing file descriptor \codeword{137}. This is a common technique used in reverse shells to redirect input/output to a network socket.
    \item \codeword{exec 137<>/dev/tcp/\textbf{ip\_addr}/\textbf{port}}: This is creating a new file descriptor \codeword{137} and opening a TCP connection to the IP address \codeword{\textbf{ip\_addr}} on port \codeword{\textbf{port}}. This is the reverse shell connection.
    \item \codeword{sh <\&137 >\&137 2>\&137}: This is executing the shell command and redirecting input, output, and errors to file descriptor \codeword{137}. This is the reverse shell payload.    
\end{itemize}
\vspace{1mm}
\textbf{Overall}: The command is attempting to establish a reverse shell connection to a remote host and execute a shell command on that host. This could be used for malicious purposes such as remote access or data exfiltration.
\vspace{1mm}
\begin{tcolorbox}\centering
\textbf{Tactic:} \underline{Execution}\quad\quad\textbf{Technique:} \underline{Unix Shell}
\end{tcolorbox}
\end{tcolorbox}

\begin{figure*}[t]
    \centering

\includegraphics[width=6. in, trim=320 80 320 80, clip]{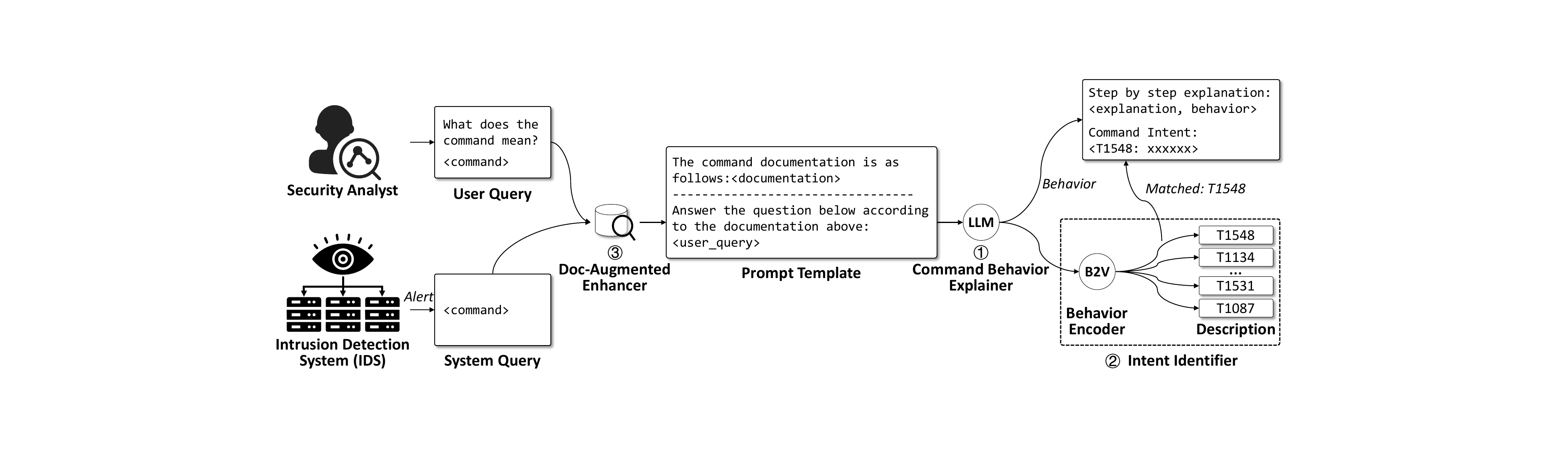}

\caption{Inference pipeline of \sys. The system receives queries from analysts and IDS. First, related documentations of the requested command are retrieved. Then a documentation-augmented prompt is created and fed into the command behavior explainer for analysis. The summarized behavior of the command is utilized for intent identification.}\label{fig:inference}
\end{figure*}

\section{System Model}
We define the system model in terms of the actors, the assumptions of their capabilities, and the security objectives.

\subsection{Actor}
In the context of security operation within an organization, we refer to the party attempting to execute malicious shell commands for assets such as sensitive information or control access as the \textit{attacker}, and the party analyzing such suspicious commands as the \textit{analyst}. In a typical security operation scenario, the malicious commands executed by the attacker are captured and flagged by the IDS for violating specific rules. Subsequently, this alert is sent to the analyst to confirm whether it constitutes a genuine attack.

\subsection{Capability}
\textit{Attacker.} We assume that the attacker can operate the shell within the system, \textit{e.g.}, they may be insiders with malicious intent or intruders who have gained initial access. They may design complex and potentially obfuscated commands to mislead the IDS and the analyst.

\textit{Analyst.} We assume that the analyst possesses basic knowledge and skills in security operation, including the ability to comprehend the semantics of shell commands with the help of natural language explanations. Furthermore, the analyst can assess whether these commands exhibit malicious or benign behavior based on their descriptions in natural language. Additionally, we assume that the analyst can decipher obfuscated commands when their obfuscation is recognized, thereby revealing the underlying undisguised commands.

\subsection{Security Objective}
The primary goal of \sys is to aid analysts in understanding the semantics and intents of shell commands, encompassing Unix Shell and PowerShell. This goal can be delineated into four key security objectives:
\begin{itemize}
\setlength{\itemsep}{5pt}
    \item \textbf{Comprehensiveness.} This denotes that the explanations provided by \sys should facilitate the analyst's comprehension of the command's semantics.

    \item \textbf{Insightfulness.} This signifies that \sys should help the analyst to determine whether the command is malicious or benign, elucidating the tactics and techniques employed if the command is indeed malicious.
    
    \item \textbf{Correctness.} This indicates that the content generated by \sys should be factually correct even when handling proprietary and private commands.
    
    \item \textbf{Portability.} This implies that \sys should function within a local environment without necessitating reliance on online services.
\end{itemize}

It is important to note that our aim is not to directly classify whether a command is malicious or not, but rather to aid the analyst in making that determination.

\begin{figure*}[t]
    \centering

\includegraphics[width=6. in, trim=300 50 310 50, clip]{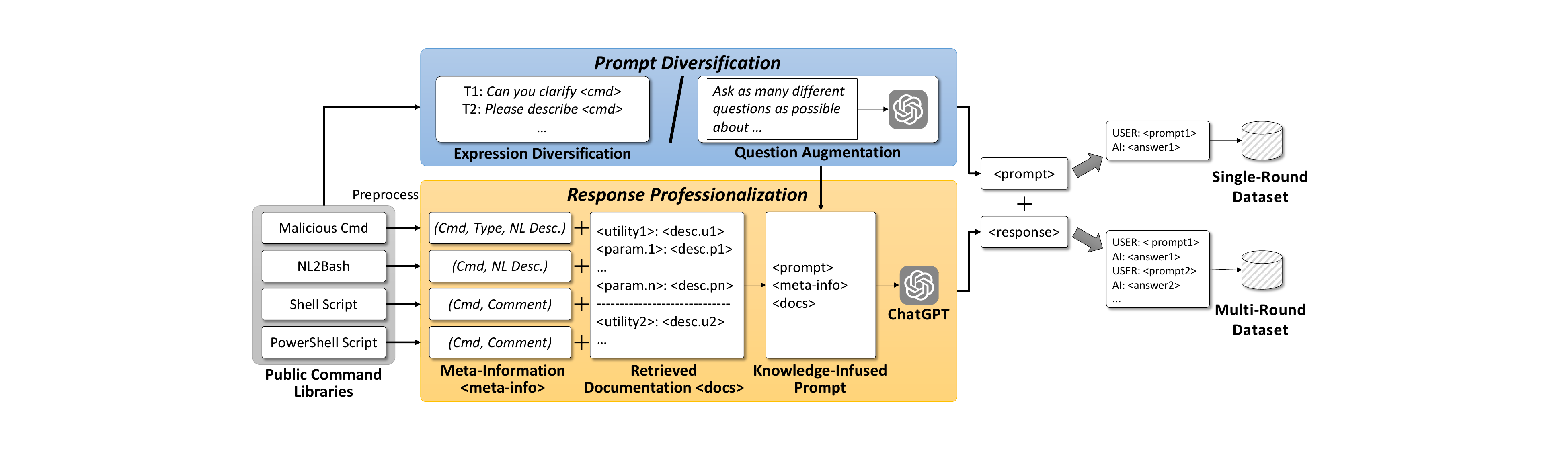}

\caption{Dataset generation pipeline of \sys. The prompt diversification module is utilized to generate prompts with diversified expressions and questions. In the response professionalization module, commands from different data sources are prompted with corresponding meta-information and command documentations for accurate responses.}\label{fig:pipeline}
\end{figure*}

\section{\sys: Detailed Construction}\label{sec:overview}

As shown in Figure~\ref{fig:inference}, \sys mainly consists of three core modules, \textit{i.e.}, behavior explainer, intent identifier, and doc-augmented enhancer.

\textbf{Behavior Explainer (\ding{192})} takes the user prompt as input and outputs the behavior explanation section (\textit{i.e.}, \textbf{Step by step explanation} \& \textbf{Overall}). The behavior explainer is obtained by fine-tuning a general LLM with a carefully-constructed dataset consisting of $\left<prompt,~response\right>$ pairs. To transform the general LLM into an expert LLM in shell command explanation, we extract knowledge from code libraries to compose high-quality responses.

\textbf{Intent Identifier (\ding{193})} takes the behavior summary from the behavior explainer and outputs the intent prediction (\textit{i.e.}, \textbf{Tactic} \& \textbf{Technique}). By mapping behavior descriptions in natural language into the standardized MITRE ATT\&CK framework, \sys assists analysts in further analyzing potential attacks.

\textbf{Doc-Augmented Enhancer (\ding{194})} improves the performance of the behavior explainer and the intent identifier by augmenting user prompt with relevant information retrieved from command documentations. To retrieve appropriate information, we design both rule-based and model-based methods to match the prompted shell command with useful texts in command documentations.

\subsection{Behavior Explainer}\label{ssec:task}

The behavior explainer gives a detailed step-by-step explanation of the command, including descriptions of the utilities, options, and parameters. The behavior explainer can point out potential malicious attempt of the command in the overall behavior summary. When a malicious attempt is specified, the behavior explainer is designed to describe the type of the malicious attempt and the countermeasures to be taken.

Existing general-purpose LLMs are not professional enough for the task of command explanation. We leverage fine-tuning to establish an expert behavior explainer with a carefully-constructed dataset of $\left<prompt,~response\right>$ pairs. Given a prompt $q_i$, the fine-tuning process aims to maximize the conditional probability of producing the response $r_i$,
\begin{equation}\label{equ:obj1}
\begin{aligned}
&\mathop{\max}\limits_{\theta}\mathop{\mathbb{E}}\limits_{\left<q_i,r_i\right>\sim \mathbbm{D}}\ p_{\theta}\left(r_i | q_i\right),
\end{aligned}
\end{equation}
\noindent where $\theta$ denotes the parameters of the behavior explainer model, initialized by a pre-trained general-purpose LLM. $\mathbbm{D}$ is the fine-tuning dataset.

A major challenge is how to construct a high-quality fine-tuning dataset. First, user prompts have diversified forms of descriptions. Therefore, we need to enrich $q_i$ to account for prompt diversity. Second, to ensure factuality of the explanation, we should infuse distilled expert knowledge into response $r_i$. We tackle these two problems by prompt diversification and response professionalization, as shown in Figure~\ref{fig:pipeline}. Specifically, we diversify the expression and augment the question of prompts in prompt diversification module. For each prompt, we infuse both meta-information (extracted from the command libraries) and retrieved related documentation to construct a knowledge-infused prompt. Subsequently, we use an online LLM service (GPT-3.5-Turbo) to organize the expert knowledge we provide and generate accurate answers/responses. Single $\left<prompt,~response\right>$ pairs constitute the single-round dialogue dataset. Multiple $\left<prompt,~response\right>$ pairs are amalgamated within one dialogue to constitute the multi-round dialogue dataset. The details of the two modules are as follows.

\subsubsection{Prompt Diversification}
We consider two aspects for prompt diversification. First, users may use different expressions to try to obtain command explanations, \textit{e.g.}, ``Please provide a detailed explanation for \codeword{<command>}.'' or ``Could you please shed some light on \codeword{<command>}?''. Second, apart from a general command explanation, users may be interested in a specific aspect of the command, \textit{e.g.}, ``What can I do with \codeword{<command>}.'' or ``What does \codeword{-c} mean in \codeword{<command>}?''

To account for different expressions, we craft a set of templates (Appendix~\ref{apx:expression}) to construct varied prompts as
\begin{equation}\label{equ:aug}
\begin{aligned}
\left<q_i,{r}_i\right> \longrightarrow \left<\mathcal{Q}(q_i, t),{r}_i\right>, t\sim \mathbbm{T},
\end{aligned}
\end{equation}
\noindent where $\mathcal{Q}(\cdot, \cdot)$ converts the original prompt $q_i$ based on the template $t$ sampled from the template set $\mathbbm{T}$.  Different expressions of the same prompt correspond to the same response $r_i$. Note that the set of templates can be easily extended using sentence rephrasing techniques in existing works~\cite{liu2023goaloriented}.

To consider other aspects of a shell command that users may be interested in, we construct an extended set of potential prompts by querying general-purpose commercial LLMs as follows.

\begin{tcolorbox}[title = {Question Augmentation Prompt}, colback=white, halign={left}]
{\small{Ask as many different questions as possible about the following command from all perspectives, and respond in the format of one question per line.}

\vspace{1mm}
Command: \texttt{\underline{<command>}}}
\end{tcolorbox}

The extended prompt set also enables us to construct $\left<multi\text{-}round~prompt,~response\right>$ pairs that allows \sys to engage in a multi-round dialogue with users as follows. 

\begin{tcolorbox}[title = {An Example of Multi-Round Dialogue}, colback=white]
{\small
\underline{USER}: What does command \codeword{bash -c '0<\&137}

\qquad\quad{\codeword{-;exec ... <\&137 >\&137 2>\&137'} do?}

\underline{\sys}:\  The command \codeword{bash} can ...  

\underline{USER}: What is the meaning of \codeword{-c}?
                                                               
\underline{\sys}:\ In \codeword{bash}, \codeword{-c} is used to ...

{... ...}

}
\end{tcolorbox}

In this way, the behavior explainer can answer questions on various aspects of the command in a multi-round interactive manner. This allows analysts to continue asking questions if there is anything unclear about the explanation.

\subsubsection{Response Professionalization}\label{ssec:accurate}
To guarantee factuality of the explanation, we construct responses by referring to professional explanations in code libraries.

\emph{Malicious shell command libraries.} We resort to three libraries that contain different kinds of malicious Unix Shell or PowerShell commands.
\begin{itemize}
\setlength{\itemsep}{2pt}
    \item \texttt{atomic-red-team.} Atomic Red Team\footnote{\url{https://github.com/redcanaryco/atomic-red-team}} is a library of tests mapped to the MITRE ATT\&CK framework. Each test includes the test command, the test name, the test description, and the attack technique it belongs to.
    \item \texttt{metta.} Metta\footnote{\url{https://github.com/uber-common/metta}} is another library of tests in the MITRE ATT\&CK framework. Each test includes the command, the name, the description, the attack technique and the tactic it belongs to. 
    \item \texttt{reverse-shell.} We generate this dataset via the Metasploit framework\footnote{\url{https://www.metasploit.com/}} as reverse shells require special attention in cybersecurity attacks. 
\end{itemize}

\emph{Benign shell command libraries.} We adopt two libraries of benign shell commands.
\begin{itemize}
\setlength{\itemsep}{2pt}
    \item \texttt{NL2Bash.} We utilize NL2Bash~\cite{Lin18NL2Bash}, which is originally designed for translating natural language descriptions into Bash commands.
    \item \texttt{The Stack.} We utilize the Unix Shell and PowerShell subsets of \texttt{The Stack}~\cite{Kocetkov22Stack}, which is a large code library. Note that, the shell scripts might contain comments, providing extra information about the commands.
\end{itemize}

The meta-information provided by the command libraries and the documentation related to each command are combined to generate accurate responses. The prompt template we use is as follows.

\begin{tcolorbox}[title = {Knowledge-Infused Prompt}, , colback=white, halign={left}]
{\small
Please refer to the command documentations and command descriptions, answer the following questions: \texttt{\underline{<prompt>}}

\vspace{2mm}
Command documentation:\\
\qquad \texttt{\underline{<docs>}}

\vspace{2mm}
Command description:\\
\qquad MITRE ATT\&CK Technique: \texttt{\underline{<meta:type>}}\\
\qquad cmd name: \texttt{\underline{<meta:name>}}\\
\qquad cmd description: \texttt{\underline{<meta:desc>}}
}
\end{tcolorbox}

\begin{figure*}[t]
    \centering

\includegraphics[width=6.in, trim=290 50 295 50, clip]{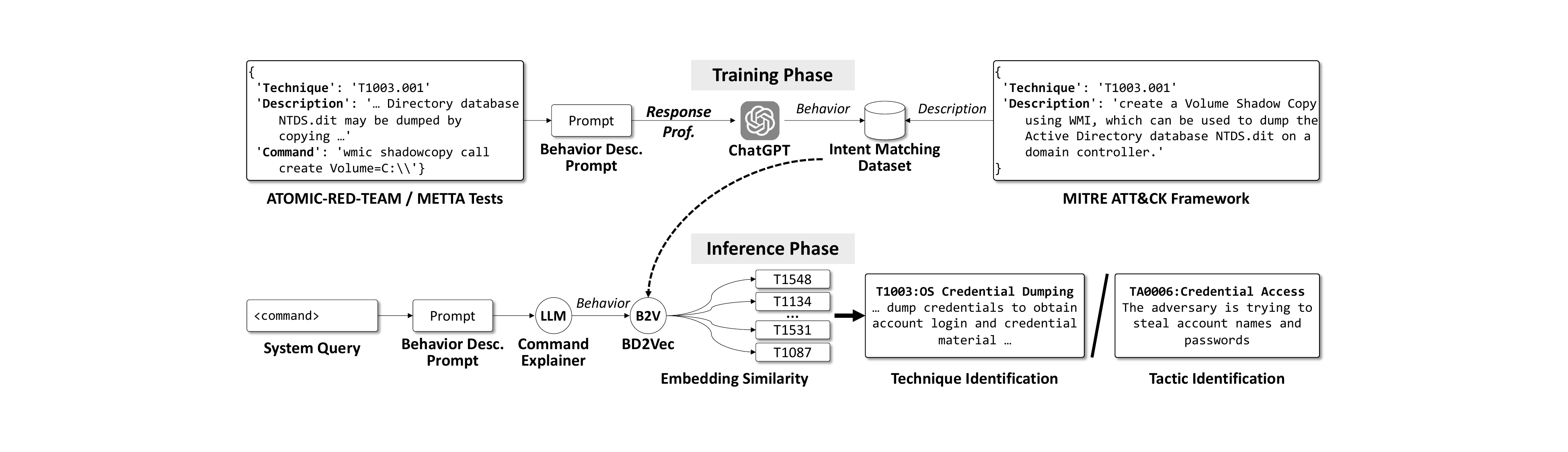}

\caption{Intent identification design of \sys. In the training phase, a Text2Vec model is fine-tuned to acquire an enhanced BD2Vec model, which maps the behavior description from LLMs and the standard description into the same embedding space. In the inference phase, the behavior description is encoded into a vector for comparison with a set of standard technique descriptions.}\label{fig:intent}
\end{figure*}

\subsection{Intent Identifier}\label{sec:intent}
In this section, we aim to identify the technique and tactic of the shell command in a standard way. First, we leverage the behavior explanation and search for the technique with the most similar description. Note that a tactic consists of a series of techniques and a technique may be used to achieve multiple tactics. Then, we predict the tactic of the command based on the identified technique.

As shown in Figure~\ref{fig:intent}, to match the behavior descriptions given by the behavior explainer and the standard technique descriptions in the MITRE ATT\&CK framework, we build a translator based on the Text2Vec model. Text2Vec is a natural language processing technique used to acquire vector representations of texts. Text2Vec models can map two texts with different syntax but similar semantics to vectors that are in close proximity. Conversely, texts with different semantics will be mapped to vectors that are distantly positioned from each other. Our intuition is to map the behavior description and the technique description into the same embedding space and then compute the distance between the embedding vectors. However, there may be a gap between the generated behavior description and the standard technique description that cannot be well bridged by a vanilla Text2Vec model. To tackle this problem, we fine-tune a Text2Vec model on a dataset of $\left<behavior,~technique~description,~label\right>$ triples to obtain an enhanced behavior-and-description-to-vector (BD2Vec) model. $label\in\{0,1\}$ denotes whether or not the $behavior$ matches the $technique~description$. More specifically, for each command sourced from \texttt{atomic-red-team} and \texttt{metta}, we acquire its behavior also using response professionalization as mentioned in \S\ref{ssec:accurate}, wherein we substitute the \texttt{<prompt>} in the knowledge-infused prompt with a behavior description prompt. The behavior description prompt is tailored to 
facilitate the generation of descriptions aligned with the style of the MITRE ATT\&CK framework through in-context learning (ICL)~\cite{Min22Rethinking}, as depicted below.

\begin{tcolorbox}[title = {Behavior Description Prompt}, , colback=white, halign={left}]
{\small
Very briefly describe what adversaries attempt to do by the following shell command:\\
Command: \texttt{\underline{<command>}}

\vspace{2mm}

Example (follow the style of the following DESCRIPTION):\\
\qquad Command: \texttt{\underline{<example-cmd>}}\\
\qquad DESCRIPTION: \texttt{\underline{<example-std-desc>}}\\

}
\end{tcolorbox}

The BD2Vec model encodes the behavior description and the standard description into vectors. In this way, the technique $t_e$ of the shell command is identified by
\begin{equation}\label{equ:technique}
\begin{aligned}
t_e = \mathop{\arg\max}\limits_{i\in\mathbbm{I}}\mathcal{S}\left(\mathcal{V}(d), \mathcal{V}(s_i)\right), 
\end{aligned}
\end{equation}
\noindent where $\mathcal{V}(\cdot)$ denotes the BD2Vec model, $\mathcal{S}(\cdot, \cdot)$ is the function to compute the similarity score, and $\mathbbm{I}$ is the set of all techniques, \textit{e.g.},  196 techniques in the MITRE ATT\&CK framework. $d$ is the description of the shell command and $s_i$ is the standard description of technique $i$.

To identify the tactic $t_a$ of a shell command, we construct a vector set $\mathbbm{I}_j$ for each tactic $j\in\mathbbm{J}$, consisting of all technique description vectors belonging to that tactic. Then the tactic is identified by
\begin{equation}\label{equ:tactic}
\begin{aligned}
t_a = \mathop{\arg\max}\limits_{j\in\mathbbm{J}}\mathrm{avg}\left(\mathop{\mathrm{Top}_k}\limits_{i\in\mathbbm{I}_j}\left[\mathcal{S}\left(\mathcal{V}(d), \mathcal{V}(s_i)\right)\right]\right), 
\end{aligned}
\end{equation}
\noindent where $\mathrm{Top}_k(\cdot)$ is a function that picks the $k$ largest elements. $\mathrm{avg}(\cdot)$ is the average function.

\subsection{Doc-Augmented Enhancer}\label{sec:doc}
The user prompt alone provides limited hint for \sys. To enhance the quality of explanation, we augment the original user prompt with relevant information retrieved from command documentations as shown in Figure~\ref{fig:inference}. The user prompt and the retrieved texts are combined as a new prompt to be fed into \sys. 

The key of command documentation retrieval is to search for most relevant information to the user prompt. To achieve this goal, we match the command in the user prompt with contents in available documentations. Similar to intent identification, we fine-tune a Text2Vec model into a command-and-documentation-to-vector (CD2Vec) model with a carefully-constructed dataset of $\left<command\right.$, $doc$, $\left.label\right>$ triples, where $label$ denotes whether the command and the documentation are related.

We utilize the Linux manual pages of Bash commands to curate a dataset as follows.

\subsubsection{Extraction of utility and option descriptions}
Given the consistent structure of Linux manual pages across utilities, we utilized regular expression matching to extract the descriptions of utilities and their respective options. This method enabled us to establish the ground-truth documentation for each utility and its associated options.

\subsubsection{Generation of shell commands}
We randomly create shell commands by combining utilities and their options, which are not necessary to be real-world commands, \textit{e.g.}, \codeword{ls~~--ignore-backups} and \codeword{ls~~--color~~-t}. As we possess the knowledge regarding the utilities and options associated with each shell command, we are able to ascertain the ground-truth documentation for each command. (from Step~1).

\subsubsection{Documentation chunking}
We partition the raw documentations into chunks by certain rules, \textit{i.e.}, word count or line feeds, simulating documentation snippets.

\subsubsection{Creation of triples}
We obtain $\left<command,~doc,~label\right>$ triples by verifying whether the documentation snippet is in the ground-truth set (from Step~2).

The doc-augmented enhancer enables \sys to explain private shell commands that are not in public datasets by retrieving relevant information from private documentations. Note that a company may define new private shell commands with the same name as but different meanings from public shell commands in the training set of LLMs.
In this case, we expect \sys to explain the private shell commands mainly according to the private documentation but not on the memorized knowledge extracted from public shell commands. To realize this purpose, in the training phase, after documentation retrieval, we substitute the objects (\textit{e.g.}, utilities) in both the command and the retrieved contexts with random strings as if they are unseen by \sys. In this way, \sys can be trained to create explanations relying on the documentations faithfully.

\section{Experiment Setup}\label{sec:evaluation}

\subsection{Prototype} 
We have implemented a fully-functional prototype of \sys. 

\begin{itemize}
    \item \emph{Behavior explainer} (\ding{192}). The behavior explainer is obtained by fine-tuning ChatGLM2-6B, an open-source bilingual (Chinese-English) chat model of 6 billion parameters~\cite{Zeng23GLM, Du22GLM}. We conduct the fine-tuning process  using four NVIDIA A100 (80GB) GPUs for four days with a batch size of 16 and a learning rate of 1e-4 for 42,000 steps. We set the maximum length of queries and responses as 1,024 tokens. The total number of tokens for fine-tuning is 232 millions. 
    \item \emph{Intent identifier} (\ding{193}). The key component of the intent identifier, \emph{i.e.}, the BD2Vec that aligns the embedding space of the generated behavior explanation and the standard description of the MITRE ATT\&CK framework, is materialized with five Text2Vec models,~\textit{i.e.}, Sentence-T5$_\mathrm{large}$~\cite{ni2021sentence}, GTR-T5$_\mathrm{XL}$~\cite{ni2021large}, SGPT~\cite{muennighoff2022sgpt}, E5$_{\mathrm{large}}$~\cite{Wang22Text}, and E5$_{\mathrm{large}}$(FT). 
    \item \emph{Doc-augmented enhancer} (\ding{194}). The key component of the doc-augmented enhancer, \emph{i.e.}, the CD2Vec model that relates the command and relevant information in documentations, is obtained by fine-tuning the second version of E5$_{\mathrm{large}}$ model. E5$_{\mathrm{large}}$ is the state-of-the-art text embedding model of 330 million parameters~\cite{Wang22Text}. The embedding size is 1,024. We fine-tune the model using low-rank adaptation (LoRA)~\cite{Hu22LoRA}, a popular parameter-efficient fine-tuning method.
\end{itemize}

\subsection{Datasets}\label{subsec:dataset}
We have curated three datasets for constructing the behavior explainer, intent identifier, and doc-augmented enhancer, respectively.

\subsubsection{Dataset for command explainer} As mentioned in \S\ref{ssec:accurate}, we construct the dataset for the command explainer based on five data sources. The dataset contains a total of 254,000 samples. We split the dataset into 9:0.5:0.5 for training, validation, and testing respectively. The test set is built to fulfil the following properties.
\begin{itemize}
\setlength{\itemsep}{2pt}
    \item \textit{Diversity.} The test set includes both malicious and benign shell commands. More than 300 types of malicious Unix Shell and PowerShell commands are included.
    \item \textit{Bilingual.} The test set includes both Chinese and English queries and responses.
    \item \textit{Single-round \& multi-round.} The test set consists of both single-round and multi-round samples for command explanation. The multi-round samples enhance the user-system interaction.
    \item \textit{Inspected by human experts.} Four computer science researchers have been recruited to inspect and refine 200 of the test samples to be served as evaluation references.
\end{itemize}

More details of the test set is presented in Table~\ref{tab:testset}. Note that all commands in our test set are real-world commands. Furthermore, there is no overlap between commands in our training set and test set. Therefore, the commands in the test set can be regarded as new and unseen.

\subsubsection{Dataset for intent identifier}
We utilize the \texttt{atomic-red-team} dataset for training and validating the BC2Vec model of the intent identifier, while the \texttt{metta} dataset is used for testing. In total, \texttt{atomic-red-team} and \texttt{metta} contain malicious shell commands of {129} techniques ({293 sub-techniques}) and {14} tactics, all of which have been labelled with ground-truth techniques and tactics by security experts. We have manually verified the labels of the test set and discovered that some of the labels were out-of-date due to the continuous maintenance and updates of the MITRE ATT\&CK framework. Consequently, we have replaced these labels with the correct ones from the most recent version of the MITRE ATT\&CK framework. Note that the original test set exhibits non-uniformity. A naive model that predicts the densest techniques can achieve a Top-1 ACC of 10.7\% and a Top-5 ACC of 41.4\%. To mitigate the impact of data distribution imbalance, we create a balanced version of the test set through resampling.

\subsubsection{Dataset for doc-augmented enhancer} As mentioned in \S\ref{sec:doc}, we construct a documentation retrieval dataset with the Linux manual pages of 1,662 Bash utilities, curating around 952,000 triples. We split the dataset into 9:0.5:0.5 for training, validating, and testing the CD2Vec model of the doc-augmented enhancer.

\begin{table*}\centering
\setlength{\abovecaptionskip}{0pt}%
\setlength{\belowcaptionskip}{0pt}%

\caption{The Overall Performance of the Command Explainer.}\label{tab:overall}

\resizebox{\linewidth}{!}{
\begin{threeparttable}[t]

\footnotesize
\setlength{\tabcolsep}{0.8mm}{
\begin{tabular}{@{}l|cccccc|cccccc@{}}
\toprule
\multirow{2}{*}[-3pt]{Model} & \multicolumn{6}{c|}{Malicious Command\tnote{\textdaggerdbl}}                                                                                        & \multicolumn{6}{c}{Benign Command\tnote{\textdagger}}                                         \\ \cmidrule(l){2-13}   
                       & \multicolumn{1}{r}{ROUGE-1} & \multicolumn{1}{r}{ROUGE-2} & \multicolumn{1}{r}{ROUGE-$\ell$} & \multicolumn{1}{r}{BLEU-4} & \multicolumn{1}{r}{{METEOR}} & \multicolumn{1}{r|}{{CIDEr}} & \multicolumn{1}{r}{ROUGE-1} & \multicolumn{1}{r}{ROUGE-2} & \multicolumn{1}{r}{ROUGE-$\ell$} & \multicolumn{1}{r}{BLEU-4}    & \multicolumn{1}{r}{{METEOR}} & \multicolumn{1}{r}{{CIDEr}}\\ \midrule             
GPT-3.5-Turbo          & 48.7                        & 24.3                        & 34.5                        & 36.3                             &{39.1} &{30.2} & 54.3                         & 27.0                        & 35.2                        & 29.5                                                                  &{40.9} &{16.8} \\                      
GPT-4                  & 45.5                        & 20.3                        & 30.5                        & 40.5                             &{32.5} &{14.6} & 51.8                         & 25.8                        & 36.2                        & 34.2                                                                  &{34.5} &{12.0} \\                      
ChatGLM2-6B            & 42.5                        & 18.0                        & 26.9                        & 35.2                             &{31.5} &{10.4} & 50.5                         & 24.3                        & 32.1                        & 30.9                                                                  &{32.5} &{9.3} \\                      
\sys                   & \textbf{\small68.9}         & \textbf{\small51.5}         & \textbf{\small58.8}         & \textbf{\small59.5}              &{\textbf{\small51.1}} &{\textbf{\small128.5}} & \textbf{\small69.3}        & \textbf{\small46.1}         & \textbf{\small53.1}         & \textbf{\small48.5}                      &{\textbf{\small50.5}} &{\textbf{\small43.0}} \\ \midrule             
Increased\tnote{\textdollar} & 62.1\%                & 186.1\%                     & 118.5\%                     & 69.0\%                           &{62.2\%} &{1137.1\%} & 37.2\%                 & 89.7\%                      & 65.4\%                      & 57.0\%                                                                &{55.7\%} &{362.9\%} \\                      
cf. GPT-4\tnote{$\star$} & 151.4\%                   & 253.7\%                     & 192.8\%                     & 146.9\%                          &{157.1\%} &{880.5\%} & 133.8\%                & 178.7\%                     & 146.7\%                     & 141.8\%                                                               &{146.4\%} &{358.7\%} \\ \bottomrule          
\end{tabular}
}

\begin{tablenotes}[flushleft]
\item[\textdaggerdbl] Malicious Command consists of data from \texttt{atomic-red-team}, \texttt{metta}, and \texttt{reverse-shell}.
\item[\textdagger] Benign Command consists of data from \texttt{NL2Bash}.
\item[\textdollar] Increased: the percentage improvement of \sys over the original ChatGLM2-6B model. 
\item[$\star$] cf. GPT-4: the achieved percentage of GPT-4 performance.
\end{tablenotes}

\end{threeparttable}}
\end{table*}
\raggedbottom
\begin{table}\centering
\setlength{\abovecaptionskip}{0pt}%
\setlength{\belowcaptionskip}{0pt}%

\caption{The Overall Performance of the Command Explainer on HumanCheck Test Set.}\label{tab:overall_humancheck}

\resizebox{\linewidth}{!}{
\begin{threeparttable}[t]

\footnotesize
\setlength{\tabcolsep}{0.8mm}{
\begin{tabular}{@{}l|cccccc@{}}
\toprule
\multirow{2}{*}[-3pt]{Model} & \multicolumn{6}{c}{HumanCheck}                                                                                             \\ \cmidrule(l){2-7} 
                       & \multicolumn{1}{r}{ROUGE-1} & \multicolumn{1}{r}{ROUGE-2} & \multicolumn{1}{r}{ROUGE-$\ell$} & \multicolumn{1}{r}{BLEU-4}    & \multicolumn{1}{r}{{METEOR}} & \multicolumn{1}{r}{{CIDEr}}\\ \midrule             
GPT-3.5-Turbo          & 62.1                        & 37.5                        & 45.9                        & 48.8                             &{43.3}  &{30.3}  \\
GPT-4                  & 54.4                        & 29.1                        & 36.9                        & 38.8                             &{31.1}  &{6.8}  \\
ChatGLM2-6B            & 55.7                        & 29.4                        & 34.9                        & 38.3                             &{32.6}  &{6.8}  \\
\sys                   & \textbf{\small69.6}         & \textbf{\small47.7}         & \textbf{\small53.3}         & \textbf{\small53.6}              &{\textbf{\small48.9}}  &{\textbf{\small47.0}}  \\ \midrule
Increased\tnote{\textdollar} & 25.0\%                & 62.2\%                      & 52.7\%                      & 39.9\%                           &{49.8\%}  &{594.4\%}  \\ 
cf. GPT-4\tnote{$\star$} & 127.9\%                   & 163.9\%                     & 144.4\%                     & 138.1\%                          &{157.0\%}  &{693.5\%}  \\ \bottomrule
\end{tabular}
}

\begin{tablenotes}[flushleft]
\item[\textdollar] Increased: the percentage improvement of \sys over the original ChatGLM2-6B model. 
\item[$\star$] cf. GPT-4: the achieved percentage of GPT-4 performance.
\end{tablenotes}

\end{threeparttable}}
\end{table}
\raggedbottom
\begin{table}\centering
\setlength{\abovecaptionskip}{0pt}%
\setlength{\belowcaptionskip}{0pt}%

\caption{End-to-End Evaluation of The Performance of the Command Explainer.}\label{tab:overall_classify}

\resizebox{\linewidth}{!}{
\begin{threeparttable}[t]

\footnotesize
\setlength{\tabcolsep}{2mm}{
\begin{tabular}{@{}l|ccc@{}}
\toprule
\multirow{2}{*}[-3pt]{Model} & \multicolumn{3}{c}{Classification Metrics}                                                                                             \\ \cmidrule(l){2-4} 
                                    & \multicolumn{1}{r}{Precision (\%)} & \multicolumn{1}{r}{Recall (\%)} & \multicolumn{1}{r}{Accuracy (\%)}\\ \midrule
GPT-3.5-Turbo                       & 78.7                        & 62.6                        &72.8\\
GPT-4                               & 76.7                        & 59.8                        &70.8\\
ChatGLM2-6B                         & 77.6                        & 47.1                        &66.7\\
\sys                                & \textbf{\small83.7}         & \textbf{\small79.2}         &\textbf{\small81.8}\\ \bottomrule
\end{tabular}
}

\end{threeparttable}}
\end{table}
\raggedbottom

\subsection{Evaluation Methods} 

To comprehensively assess the performance of \sys, we conduct both quantitative and qualitative evaluations. 

\subsubsection{Quantitative evaluations}
We use the following quantifiable metrics to evaluate the performance of \sys. 

\begin{itemize}

\item \emph{Quantifiable metric for behavior explainer}. We adopt four well-established quantifiable metrics, \textit{i.e.}, ROUGE~\cite{lin04rouge}, BLEU~\cite{Papineni02Bleu}, METEOR~\cite{BanerjeeL05} and CIDEr~\cite{VedantamZP15} to evaluate the performance of the behavior explainer, which provide token-level interpretable evaluations of natural language outputs. METEOR and CIDEr are two advanced metrics that have demonstrated a high correlation with human judgments~\cite{BanerjeeL05,VedantamZP15}.

\begin{itemize}
\setlength{\itemsep}{5pt}
    \item ROUGE (Recall-Oriented Understudy for Gisting Evaluation)~\cite{lin04rouge} is an NLP metric that compares a machine-produced text against the ground-truth text. A higher ROUGE-\textit{k} means a larger \textit{k}-grams overlapping between the generated text and the ground-truth text. \textit{k} is usually set as 1~or~2. ROUGE-$\ell$ measures the longest common sub-sequence (LCS) of the generated text and the ground-truth text.
    
    \item BLEU (Bilingual Evaluation Understudy)~\cite{Papineni02Bleu} is an NLP metric that is precision-oriented. A higher BLEU-\textit{k} indicates a larger \textit{k}-grams overlapping between the generated text and the ground-truth text. \textit{k} is usually set to 4. 
    \item METEOR (Metric for Evaluation of Translation with Explicit ORdering)~\cite{BanerjeeL05} is an advanced NLP metric that takes into account word order, stemming matching and synonymy matching, aspects which are not considered by BLEU.
    
    \item CIDEr (Consensus-Based Image Description Evaluation)~\cite{VedantamZP15} is an advanced NLP metric originally designed to compute the similarity between a machine-generated image description and the ground-truth description.
    
\end{itemize}

We also conduct end-to-end evaluation on the performance of the behavior explainer by using the generated command explanations for command classification, simulating the scenario when security analysts are presented with the command explanations. In this way, accuracy, precision, and recall are utilized as metrics.
    
\item \emph{Quantifiable metric for intent identifier}. We use Top-\textit{k} ACC to assess the accuracy of the intent identifier in  detecting the techniques and tactics of malicious commands. Top-\textit{k} ACC represents the proportion of cases where the ground-truth label is among the top $k$ labels predicted (ranked by similarity scores). A higher Top-\textit{k} ACC means that the matching method can better infer the attacker's intent.

\item \emph{Quantifiable metric for doc-augmented enhancer}. AUC-ROC is used to evaluate the performance of the doc-augmented enhancer. A higher AUC-ROC (Area Under the ROC curve) score indicates that the CD2Vec model can effectively encode related command-documentation pairs to be in close proximity, while distinctly separating unrelated pairs.

\end{itemize}

\subsubsection{Qualitative evaluations}
 We have recruited {52} undergraduate and graduate students majoring in computer science to assess the responses of \sys. The detailed evaluation process and results are provided in \S\ref{ssec:user}.

\subsection{Baselines} 

\subsubsection{Baselines for behavior explainer} For behavior explanation, we compare \sys with GPT-3.5-Turbo~\cite{Brown20Language, Ouyang22Training}, GPT-4~\cite{OpenAI23GPT4} and the original ChatGLM2-6B~\cite{Du22GLM, Zeng23GLM}. GPT-3.5-Turbo and GPT-4 are two commercial state-of-the-art LLMs and ChatGLM2-6B is an open-source bilingual (Chinese-English) model.

\subsubsection{Baselines for intent identifier} For intent identifier, we compare \sys with GPT-3.5-Turbo and GPT-4. In particular, we demonstrate a chain-of-thought~\cite{Wei22Chain} process in the prompt together with an example of a command and its corresponding technique. Then, we query GPT-3.5-Turbo and GPT-4 to identify a given command's technique and tactic.

\subsubsection{Baselines for doc-augmented enhancer} For doc-augmented enhancer, we compare \sys with four state-of-the-art embedding models, \textit{i.e.}, Sentence-T5$_\mathrm{large}$~\cite{ni2021sentence}, GTR-T5$_\mathrm{XL}$~\cite{ni2021large}, SGPT~\cite{muennighoff2022sgpt}, and E5$_{\mathrm{large}}$~\cite{Wang22Text}, which are state-of-the-art Text2Vec models on the massive text embedding benchmark~\cite{Muennighoff23MTEB}. Detailed information of the baseline models is summarized in Table~\ref{tab:generative_models} and \ref{tab:text2vec_models}.

\section{Evaluation Results}\label{sec:result}

\begin{table}[tt]\centering
\setlength{\abovecaptionskip}{0pt}%
\setlength{\belowcaptionskip}{0pt}%
\caption{The Performance on Three Tasks.}\label{tab:evaltask}

\resizebox{\linewidth}{!}{
\begin{threeparttable}

\footnotesize
\setlength{\tabcolsep}{2mm}{
\begin{tabular}{@{}l|cccc@{}}
\toprule
\multirow{2}{*}[-3pt]{Model}    & \multicolumn{4}{c}{Explanation \& Explanation w/ Doc.\tnote{\textdagger}} \\ \cmidrule{2-5}
                                & \multicolumn{1}{r}{ROUGE-1}   & \multicolumn{1}{r}{ROUGE-2} & \multicolumn{1}{r}{ROUGE-$\ell$} & \multicolumn{1}{r}{BLEU-4}    \\ \midrule
GPT-3.5-Turbo                   & 49.3                          & 24.0                        & 31.4                             & 28.8                           \\
GPT-4                           & 48.5                          & 23.7                        & 32.7                             & 36.9                           \\
ChatGLM2-6B                     & 46.3                          & 21.9                        & 28.9                             & 31.6                           \\
\sys                            & \textbf{\small68.4}           & \textbf{\small48.1}         & \textbf{\small54.6}              & \textbf{\small52.9}            \\ \midrule
Increased\tnote{\textdollar}    & 47.7\%                        & 119.6\%                     & 88.9\%                           & 67.4\%                           \\ 
cf. GPT-4\tnote{$\star$}        & 141.0\%                       & 203.0\%                     & 167.0\%                          & 143.3\%                          \\ \midrule\midrule
\multirow{2}{*}[-3pt]{Model}    & \multicolumn{4}{c}{Behavior Summarization} \\ \cmidrule{2-5}
                                & \multicolumn{1}{r}{ROUGE-1}   & \multicolumn{1}{r}{ROUGE-2} & \multicolumn{1}{r}{ROUGE-$\ell$} & \multicolumn{1}{r}{BLEU-4}    \\\midrule
GPT-3.5-Turbo                   & 51.1                          & 26.0                        & 39.5                        & 43.8                                \\
GPT-4                           & 44.8                          & 18.2                        & 30.5                        & 42.7                                \\
ChatGLM2-6B                     & 41.6                          & 15.7                        & 26.8                        & 38.5                                \\
\sys                            & \textbf{\small70.3}           & \textbf{\small53.9}         & \textbf{\small62.1}         & \textbf{\small63.8}                 \\\midrule
Increased\tnote{\textdollar}    & 69.0\%                        & 243.3\%                     & 131.7\%                     & 65.7\%                           \\ 
cf. GPT-4\tnote{$\star$}        & 156.9\%                       & 296.2\%                     & 203.6\%                     & 149.4\%                          \\ \bottomrule
\end{tabular}
}

\begin{tablenotes}[flushleft]
\item[\textdagger] Explanation \& Explanation w/ Doc. include both the explanation task and the explanation with documentation task.
\item[\textdollar] Increased: the percentage improvement of \sys over the original ChatGLM2-6B model. 
\item[$\star$] cf. GPT-4: the achieved percentage of GPT-4 performance.
\end{tablenotes}

\end{threeparttable}}
\end{table}

\subsection{Command Explanation}
\subsubsection{Overall Performance}\label{ssec:overall}
As shown in Table~\ref{tab:overall}, we compare the explanation capability of \sys with baselines regarding both malicious and benign commands. We can see that \sys achieves the best explanation performance on both malicious and benign commands in terms of all six metrics, improving the vanilla ChatGLM2-6B by 37.2\%$\sim$1137.1\%. Note that, the three baselines perform worse on malicious commands than on benign commands in terms of ROUGE-\textit{k}, which means some expected information related to the malicious commands is not mentioned in their responses. In contrast, \sys shows equal and even better explanation performance on malicious commands, \textit{e.g.,} \sys outperforms the second-best by 41\% and 32\% in terms of ROUGE-$\ell$ and BLEU-4, respectively. This may result from the response professionalization that we design in \S\ref{ssec:accurate} to incorporate knowledge of malicious commands into \sys. We also evaluate \sys and the three baselines on the HumanCheck test set, which is inspected and corrected by four computer science researchers. We can see in Table~\ref{tab:overall_humancheck} that \sys also achieves the best performance on HumanCheck test set.

\subsubsection{End-to-End Evaluation of Overall Performance} In order to evaluate the insightfulness of the command explanations for aiding security analysts in making judgments, we provide the command explanations to GPT-3.5-Turbo for command classification. As depicted in Table~\ref{tab:overall_classify}, \sys achieves the highest precision (83.7\%), recall (79.2\%) and accuracy (81.8\%). 
It demonstrates that the explanations generated by \sys are more effective in facilitating command comprehension for identifying malicious commands.

\begin{figure}[tt]
\setlength{\subfigcapskip}{0pt}
	\centering  
        \subfigure[\small Different top-\textit{p} for inference (temperature=0.8)]{   
    		\includegraphics[width=.8\linewidth, trim=0 5 -10 0, clip]{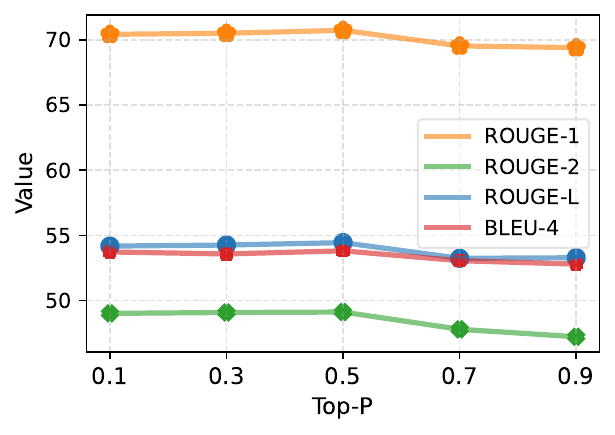} 
        }
        \\
    	\subfigure[\small Different temperature for inference (top-\textit{p}=0.8)]{   
    		\includegraphics[width=.8\linewidth, trim=0 5 -10 0, clip]{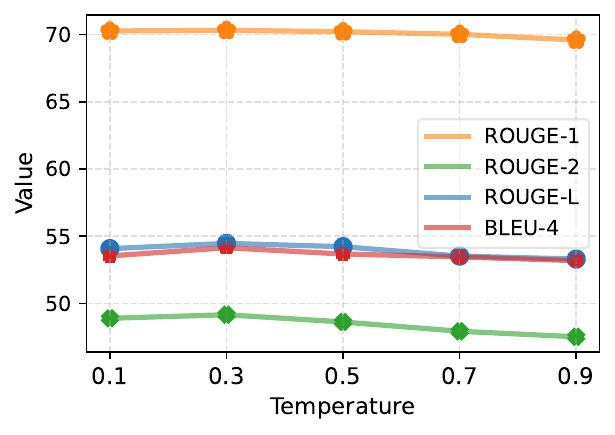} 
    	}
	\caption{The impact of temperature and top-\textit{p} on the performance of \sys.}    
	\label{fig:tempature_topp}    
\end{figure}

\subsubsection{Different Tasks}
We compare \sys with the baselines on two tasks, \textit{i.e.}, step-by-step explanation and behavior summarization. We present the results in Table~\ref{tab:evaltask}. As shown in Table~\ref{tab:evaltask}, \sys outperforms the three baselines on the step-by-step explanation and behavior summarization tasks by a large margin (\textit{e.g.,} over 40\% and 30\% in terms of ROUGE-$\ell$ and BLEU-4, respectively).

\subsubsection{Different Expressions of Query}
Since users might use different ways to ask \sys to explain a command, in this part, we evaluate \sys's robustness against diversified expressions of explanation queries. We present the evaluation results of three types of query expressions in Table~\ref{tab:diversified} in Appendix~\ref{apx:diversified}, including the original query, the diversified query and the diversified query with documentations. We create diversified queries by converting the original queries into different expressions using Equation~\eqref{equ:aug}.
We can see that \sys performs consistently on all three query expressions and achieves the best performance comparing with the baselines. This suggests that \sys can generate comprehensive explanations without explicit or intentional prompting. Moreover, \sys demonstrates the best performance on diversified queries with documentation. This underscores the robustness of \sys to query expressions and its proficiency in leveraging command documentations. We notice that the performance of the baselines degrades when queries are in different expressions. It is because existing general-purpose LLMs are not trained to account for the variance in queries. In contrast, \sys provides complete analysis when requested by different expressions of explanation queries. 

\subsubsection{Different Languages}
We also evaluate the bilingual capability of \sys comparing with the three baselines. As shown in Table~\ref{tab:language} in Appendix~\ref{apx:language}, we test four models on the same tasks both in Chinese and English. \sys achieves the best performance in both languages. Note that it seems that these four models have better performance in English than in Chinese, however, the metrics for Chinese and English are computed in different ways in terms of word segmentation (\textit{a.k.a.}, \textit{tokenization}) and thus are not comparable.

\subsubsection{Different Inference Hyper-Parameters}
In this part, we evaluate the impact of the inference hyper-parameters, including the temperature and the top-\textit{p} parameters. We set temperature and top-\textit{p} from 0.1 to 0.9, evaluate \sys on the HumanCheck test set, and present the evaluation results in Figure~\ref{fig:tempature_topp}. We find that these two parameters have little impact on the performance of \sys. When we vary the temperature, the performance of \sys only changes by 1\%$\sim$3\%, and reaches the best result when temperature=0.3. When we vary \textit{p}, the performance of \sys only changes by 2\%$\sim$4\%, and achieves the best result when \textit{p}=0.5.

\begin{figure}[t]
    \centering
\includegraphics[width=.8\linewidth, trim=0 0 0 0, clip]{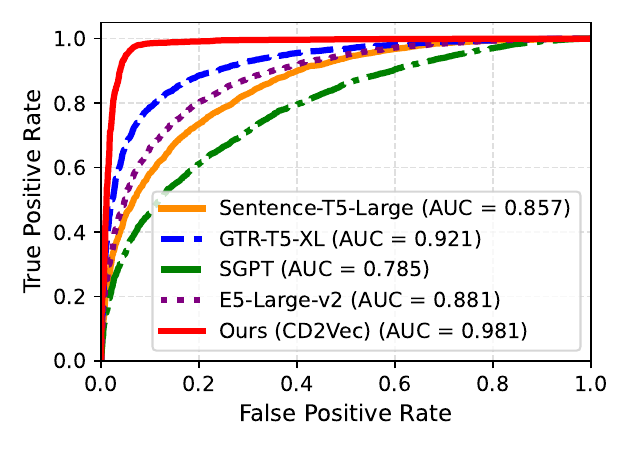}

\caption{ROC curves of five Text2Vec models. Our CD2Vec model obtains the highest AUC, indicating the best discriminating ability.}\label{fig:auc_roc}
\end{figure}

\subsection{Intent Identification}\label{ssec:intent}

We compare the intent identification capability of \sys with baselines GPT-3.5-Turbo and GPT-4. We present the evaluation results for the original test set in Table~\ref{tab:test_intent} and those for the balanced test set in Table~\ref{tab:test_intent_balanced}. Overall, the fine-tuned version of E5$_\mathrm{large}$ achieves the best performance on both technique and tactic identification. All five of our materializations of the intent identifier show superior performance compared to GPT-3.5-Turbo and GPT-4. For example, SGPT, though performing the worst among our five materializations of the intent identifier, still surpasses GPT-3.5-Turbo and GPT-4 by 16.4\% and 7.1\% in terms of Top-1 ACC on technique and tactic identification, respectively. We attribute this to our behavior-oriented intent identification method. Note that, the Top-1 and Top-5 ACC of \sys (52.4\% and 83.0\%) significantly surpass those of the naive model that predicts the densest techniques mentioned in \S\ref{subsec:dataset}, \textit{i.e.}, 10.7\% and 41.4\%. Additionally, we observe only minor differences between the results of Table~\ref{tab:test_intent} and Table~\ref{tab:test_intent_balanced}. This demonstrates that the superiority of \sys is not contingent upon data distribution imbalance.

\subsection{Doc-Augmented Enhancer}\label{ssec:auc_roc_exp}
A Text2Vec model attributing high similarity scores to related command-documentation pairs and low scores to unrelated pairs is more proficient in retrieving the correct documentation snippets within a documentation database.
From the dataset detailed in \S\ref{sec:doc}, we randomly select 9,782 $\left<command,~doc,~label\right>$ triples, both related and unrelated, as our test set. We compare our CD2Vec with four baselines. The ROC curves are presented in Figure~\ref{fig:auc_roc}, where our CD2Vec achieves the highest AUC of 0.981. This performance significantly surpasses that of the baselines, of which the average AUC is 0.858. Note that the baseline models each have a larger number of model parameters than ours. This result highlights that our lightweight doc-augmented enhancer is able to determine the relevance between a given shell command and a documentation snippet.




\begin{table}\centering

\setlength{\abovecaptionskip}{0pt}%
\setlength{\belowcaptionskip}{10pt}%
\caption{Technique and Tactic Identification Performance on the Original Test Set.}\label{tab:test_intent}

\resizebox{\linewidth}{!}{
\begin{threeparttable}[t]
\renewcommand{\arraystretch}{1}

\footnotesize
\setlength{\tabcolsep}{2.2mm}{
\begin{tabular}{@{}l|ccc|c@{}}
\toprule
\multirow{2}{*}[-2pt]{Model} & \multicolumn{3}{c|}{Technique (ACC)} & \multicolumn{1}{c}{Tactic (ACC)}
                    \\ \cmidrule(l){2-5} 
                   & \multicolumn{1}{c}{Top-1} & \multicolumn{1}{c}{Top-5} & \multicolumn{1}{c|}{Top-10} & 
                   \multicolumn{1}{c}{Top-1} \\ \midrule
GPT-3.5-Turbo & 25.6     &   33.8     &   35.0     &   45.7 \\
GPT-4 & 26.0     &   32.5     &   36.2     &   55.5 \\\midrule
Sentence-T5$_{\mathrm{large}}$          & 45.4                       &  80.1  &  87.3   & 69.4   \\
GTR-T5$_{\mathrm{XL}}$                  & 50.8                        &  79.6  &  87.5   & 70.5  \\
SGPT            &   42.4    &   75.0   &   84.3   &   62.6                  \\
E5$_{\mathrm{large}}$      &  48.5      &     78.2     &  87.1    &    69.3                 \\
E5$_{\mathrm{large}}$ (FT)     & \textbf{\small52.4}     &      \textbf{\small83.0}     & \textbf{\small90.4}     &  \textbf{\small75.0}  \\ \bottomrule
\end{tabular}
}

\end{threeparttable}}
\end{table}
\raggedbottom

\begin{table}\centering

\setlength{\abovecaptionskip}{0pt}%
\setlength{\belowcaptionskip}{10pt}%
\caption{Technique and Tactic Identification Performance on the Balanced Test Set.}\label{tab:test_intent_balanced}

\resizebox{\linewidth}{!}{
\begin{threeparttable}[t]
\renewcommand{\arraystretch}{1}

\footnotesize
\setlength{\tabcolsep}{2.2mm}{
\begin{tabular}{@{}l|ccc|c@{}}
\toprule
\multirow{2}{*}[-2pt]{Model} & \multicolumn{3}{c|}{Technique (ACC)} & \multicolumn{1}{c}{Tactic (ACC)}
                    \\ \cmidrule(l){2-5} 
                   & \multicolumn{1}{c}{Top-1} & \multicolumn{1}{c}{Top-5} & \multicolumn{1}{c|}{Top-10} & 
                   \multicolumn{1}{c}{Top-1} \\ \midrule
GPT-3.5-Turbo & 27.0     &   32.9     &   34.1     &   43.3 \\
GPT-4 & 28.1     &   32.0     &   34.9     &   52.3 \\\midrule
Sentence-T5$_{\mathrm{large}}$          & 51.2                       &  78.8  &  86.6   & 66.7   \\
GTR-T5$_{\mathrm{XL}}$                  & 51.3                        &  80.2  &  88.1   & 69.4  \\
SGPT            &   48.5    &   78.4   &   86.8   &   67.3                  \\
E5$_{\mathrm{large}}$      &  50.5      &     81.3     &  87.8    &    69.0                 \\
E5$_{\mathrm{large}}$ (FT)     & \textbf{\small56.0}     &      \textbf{\small82.1}     & \textbf{\small89.3}     &  \textbf{\small74.3}  \\ \bottomrule
\end{tabular}
}

\end{threeparttable}}
\end{table}
\raggedbottom

\begin{figure*}[tt]
\setlength{\subfigcapskip}{0pt}
\centering  
\subfigure[\small Details of commands]{   
\includegraphics[width=.3\linewidth, trim=0 0 -0 20, clip]{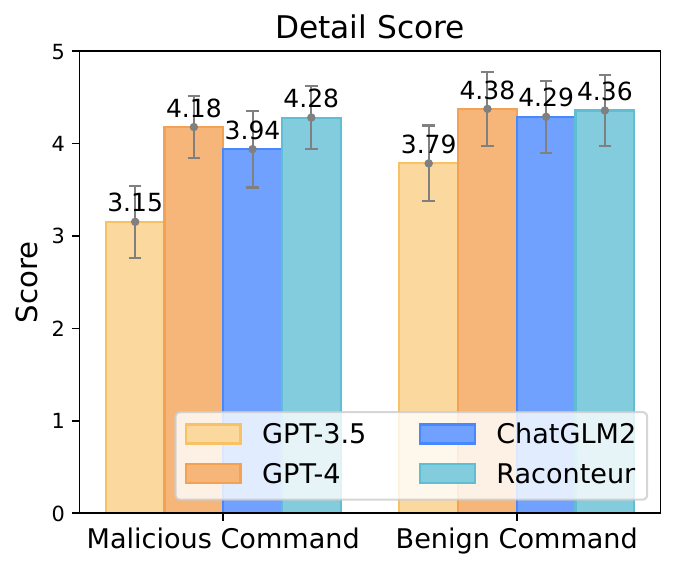}\label{fig:detail} %
}
\hfill
\subfigure[\small Intents of commands]{   
\includegraphics[width=.3\linewidth, trim=0 0 0 20, clip]{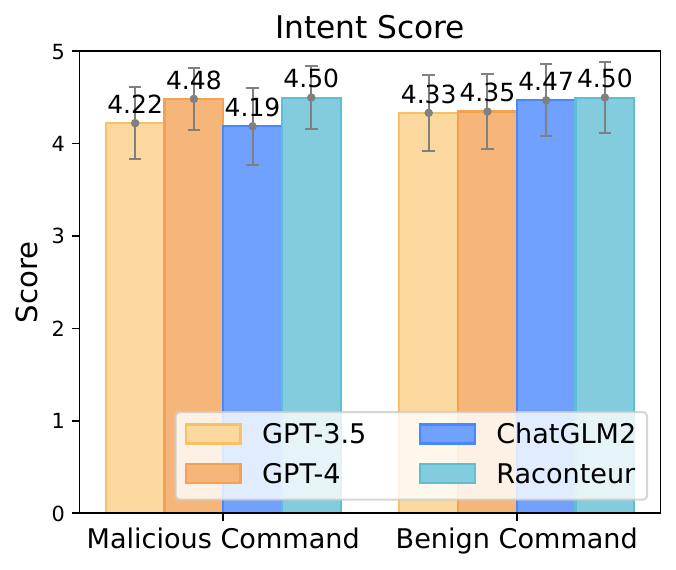}\label{fig:user_intent}
}
\hfill
\subfigure[\small Malicious or benign]{   
\includegraphics[width=.3\linewidth, trim=0 0 0 20, clip]{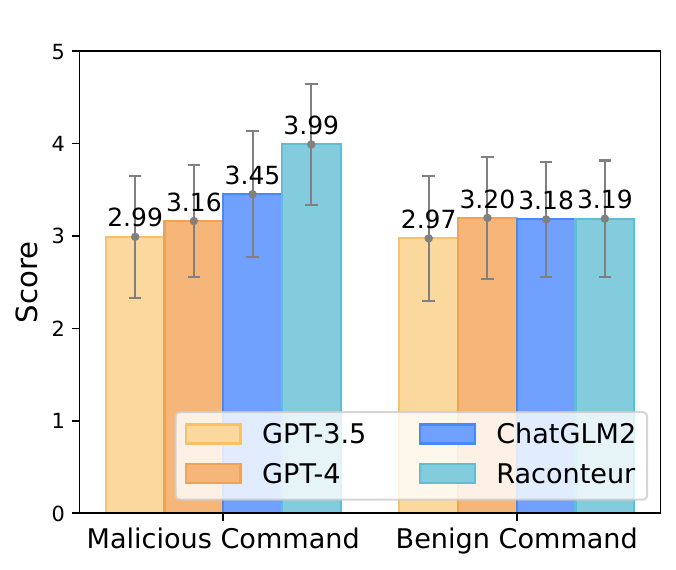}\label{fig:malicious}
}
\\\vspace{2mm}
\subfigure[\small Correctness]{   
\includegraphics[height=.25\linewidth, trim=0 0 0 20, clip]{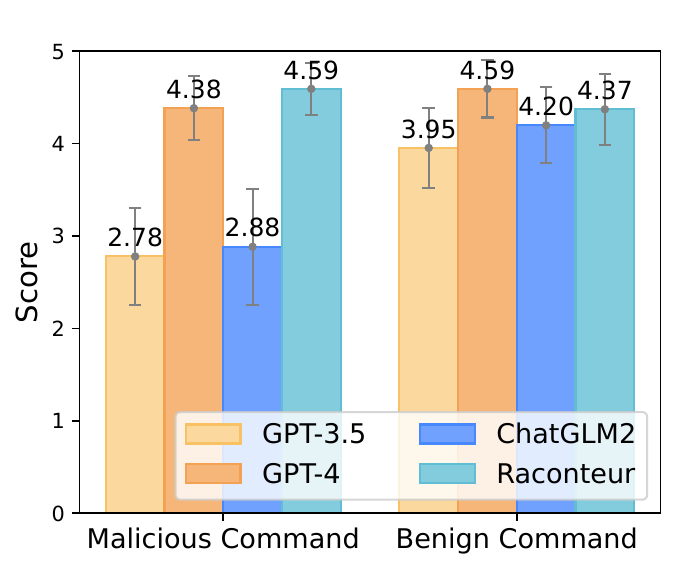}\label{fig:correctness}
}
\hspace{1cm}
\subfigure[\small Preference]{   
\includegraphics[height=.25\linewidth, trim=0 0 0 0, clip]{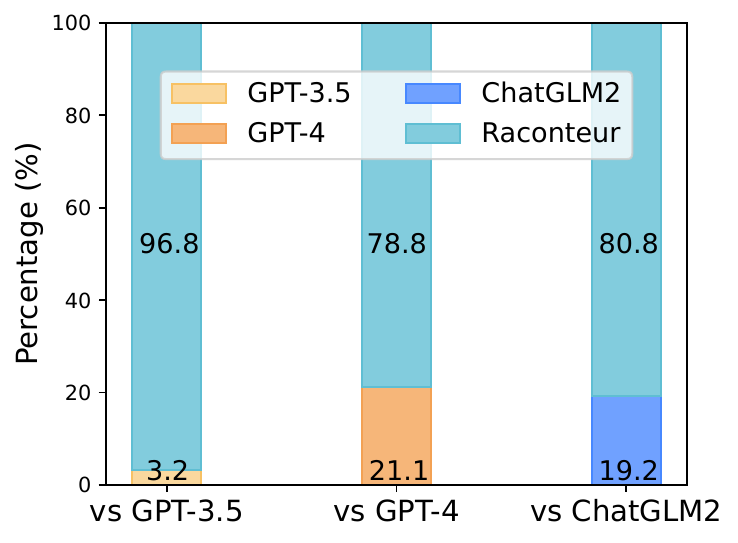}\label{fig:preference}
}
\caption{The results of the user study. \sys achieves the best comprehensiveness and correctness for malicious commands and the second-best for benign commands (slightly lower than GPT-4) (a \& d). \sys demonstrates superior insightfulness on both malicious and benign commands (b \& c). Participants favor the explanations provided by \sys over those of baselines (e).}    
\label{fig:user_part1}    
\end{figure*}

\subsection{User Study}\label{ssec:user}
We conduct a user study to qualitatively evaluate the performance of \sys.
We have recruited {52} undergraduate and graduate students majoring in computer science to answer three sets of questions. The participants are asked to evaluate their proficiency in shell command, \textit{i.e.}, at elementary, intermediate, and advanced levels\footnote{In general, advanced level indicates more than 7 years of Unix Shell or PowerShell experience, intermediate is 3$\sim$7 years, and elementary is 0$\sim$3 years.}. The proportions of elementary, intermediate and advanced participants are {35\%, 38\% and 27\%}, respectively. In the first part, each question consists of one command and the corresponding explanation (from three baselines or \sys). Participants are requested to provide three scores, \textit{i.e.}, to rate the explanation based on whether it helps them understand the details and intent of the command, as well as whether it helps them determine whether the command is malicious or not, using a scale of 1$\sim$5 points. In the second part, participants are asked to compare the reference answer and the response from LLMs, and rate the correctness of the response on a scale of 1$\sim$5 points. The first two parts consist of 40 questions, with 20 malicious commands and 20 benign commands, along with responses from \sys, GPT-4, GPT-3.5-Turbo, and ChatGLM2. In the third part, each question consists of one command and two responses from different methods. Participants are asked to choose the one they prefer and after that describe the reason why a specific response is considered good. This part includes 12 questions comparing \sys to GPT-4, GPT-3.5-Turbo, and ChatGLM2.

\subsubsection{Comprehensiveness \& Correctness} Comprehensiveness indicates that the explanation helps individuals understand the details of the command. As depicted in Figure~\ref{fig:detail}, \sys achieves the highest score for malicious commands and the second-highest score for benign commands (slightly lower than GPT-4), which aligns with the results in \S\ref{ssec:overall}. This demonstrates that \sys can generate comprehensive explanations, aiding the understanding of both malicious and benign commands. As shown in Figure~\ref{fig:correctness}, \sys achieves the best correctness on malicious commands and the second-best on benign commands. We notice that GPT-3.5-Turbo gets relatively unsatisfying results. We inspect the responses of GPT-3.5-Turbo and find that it tends to return brief and short explanations when not explicitly prompted.

\subsubsection{Insightfulness} Insightfulness indicates that the explanation helps individuals understand the intent of the command, enabling them to determine whether the command is malicious or benign. As depicted in Figure~\ref{fig:user_intent} and Figure~\ref{fig:malicious}, \sys demonstrates superior performance on both malicious and benign commands. It is noteworthy that \sys outperforms baselines in assisting individuals to identify malicious commands.


\subsubsection{Preference} Preference directly represents the likability of \sys compared with those of baselines. As shown in Figure~\ref{fig:preference}, participants favor the explanations provided by \sys over those of baselines. This validates the high quality of explanations generated by \sys.

Based on the free descriptions regarding the reasons for preference, we identify two factors that are deemed beneficial and valuable for command analysis, \textit{i.e.,} detailed formatted explanations and explicit warnings. It shows that more detailed and formatted responses with a clear hierarchical structure are preferred. Furthermore, explicit warnings about the malicious nature of commands can enhance analysts' alertness.

\section{Related Work}\label{sec:relatedwork}
In this section, we briefly review the most related works on human-oriented code explanation and machine-oriented log analysis.

\subsection{Human-Oriented Code Explanation}
\textbf{Code Explanation.} 
The category of works within code explanation endeavors to generate comprehensive explanations that aid in the better understanding of code by humans. Over the recent years, the emergence of large-scale pre-trained models has catalyzed several initiatives in the field of code explanation. Notably,
Sarse~\textit{et~al.}~\cite{Sarsa22Automatic} and MacNeil~\textit{et~al.}~\cite{macneil23experiences, MacNeil22Generating} explored and evaluated generating code explanations by large language models (LLMs) for education purposes. Denny~\textit{et~al.}~\cite{Denny22Robosourcing} introduced and evaluated robosourcing for educational resources, \textit{i.e.}, querying LLMs to replace some of the work traditionally performed by the crowd. Leinonen~\textit{et~al.}~\cite{Leinonen23comparing} compared the explanation generated by LLMs with those created by students, and found that LLMs outperformed students in terms of accuracy and understandability. 

Existing works above~\cite{Sarsa22Automatic, macneil23experiences, MacNeil22Generating, Denny22Robosourcing, Leinonen23comparing} utilized cloud services of closed-source LLMs (\textit{e.g.}, OpenAI's Codex~\cite{Chen21Codex} and GPT-3~\cite{Brown20Language}) for normal code explanation. However, the use of cloud services is restricted when it comes to explaining proprietary commands, due to their sensitive nature and close association with company operations and internal network states. In contrast, \sys is founded upon a customized local LLM and incorporates support for handling malicious commands, thereby enhancing its capability to assist in shell log auditing. This unique approach not only ensures a more secure and tailored solution for code explanation, especially in scenarios involving proprietary commands, but also aligns with the imperative need for heightened security measures in contemporary software development practices.

\textbf{Code Summarization.} The objective of code summarization is to generate natural language comments for code, which aligns with the goal of code explanation to a certain extent. Early endeavors in code summarization were information retrieval-based. Haiduc~\textit{et~al.}~\cite{Haiduc10On}, Eddy~\textit{et~al.}~\cite{Eddy13Evaluating} and Wong~\textit{et~al.}~\cite{Wong15CloCom} proposed to search comments from similar code snippets to create summaries. Moreno~\textit{et~al.}~\cite{Moreno13Automatic} extracted keywords in code snippets for Java class summaries. However, these information retrieval-based methods falter when no similar code snippet exists or when identifiers in the code are poorly named, a common occurrence in the context of malicious shell commands.

In recent years, learning-based methods have gained traction. Allamanis~\textit{et~al.}~\cite{Allamanis16Convolutional} proposed a convolutional attention network for extreme (very short) code summarization. Iyer~\textit{et~al.}~\cite{Iyer16Summarizing} used long short-term memory (LSTM) networks with attention to produce summaries of C\# and SQL code snippets. Hu~\textit{et~al.}~\cite{Hu18Deep} and Wan~\textit{et~al.}~\cite{Wan18Improving} incorporated an abstract syntax tree structure as well as sequential content of code snippets into an encoder-decoder model for better comment generation. Wei~\textit{et~al.}~\cite{Wei19Code} proposed a joint model to take advantage of the dulity of code generation and code summarization. Given the success of transformer in NLP, Wang~\textit{et~al.}~\cite{Wang20TranS} and Clement~\textit{et~al.}~\cite{Clement20PyMT5} applied transformer in code-to-text (code summarization) and text-to-code (code generation) tasks. 

Considering the promising effectiveness of graph neural networks (GNNs) on structured data, Fernandes~\textit{et~al.}~\cite{Fernandes19Structured}, Wang~\textit{et~al.}~\cite{Wang21CoCoSum} and Wang~\textit{et~al.}~\cite{Wang22GypSum} utilized GNN to reason about the long-distance relationships in code. Stepping into the era of pre-training language models (PLMs), a series of multi-tasking PLMs \cite{Lu21CodeXGLUE, Feng20CodeBERT, Wang21CodeT5, Guo21GraphCodeBERT, Wang23CodeT5p, Li23StarCoder} for code have been proposed for various code understanding and generation tasks, including code summarization. 

While code summarization can be viewed as a form of brief explanation, detailed explanations are preferred in practice for better comprehension. Moreover, existing methods are developed for general code, whereas \sys supports explaining proprietary commands based on documentation and analyzing malicious commands at a high-level through technique and tactic identification.

\subsection{Machine-Oriented Log Analysis}
In the realm of cybersecurity, there has been a significant focus on the automated analysis of logs for various downstream tasks, with an emphasis on providing explanations tailored for machine understanding rather than human consumption. Notably, Crespi~\textit{et~al.}~\cite{Crespi21Identifying} applied unsupervised NLP methods on honeypot command logs to cluster IP addresses aiming at botnet detection. Boffa~\textit{et~al.}~\cite{Boffa22Towards} leveraged bag of words and Word2Vec to learn representations from honeypot logs and identify groups of similar SSH/Telnet sessions and attacks. These endeavors predominantly represent data mining efforts aimed at offering initial insights derived from log data.

Expanding on this, Boffa~\textit{et~al.}~\cite{Boffa23LogPr} took a step further by delving into the identification of specific tactics employed by attackers within individual segments of shell sessions. While these efforts mark significant progress, they primarily yield trace-based or label-like outcomes. In contrast, our proposed system goes beyond mere traces or labels, offering comprehensive and easily comprehensible natural language explanations for commands. Moreover, \sys not only identifies the tactics employed by attackers but also discerns more granular techniques from individual commands, rather than analyzing entire sessions.

In a somewhat different vein, other research efforts \cite{Dietm22new, Houidi22Towards} have focused on learning representations from network data for predictive tasks such as delay prediction. These pursuits, while valuable, are orthogonal to our primary objective of explicating and understanding malicious activities within command logs. Our emphasis lies in providing clear, human-readable insights into the nature of potentially harmful commands, thereby enhancing the interpretability of cybersecurity log analysis.

\section{Conclusion \& Future Work}\label{sec:conclusion}

In this work, we present the design, implementation, and evaluation of \sys, a knowledgeable, insightful and portable shell command explainer powered by LLM, which provides bottom-up explanations of shell commands and assists shell log auditing in security operation. \sys is infused with professional knowledge to provide comprehensive explanations on shell commands, including not only what the command does but also why the command does it. Our extensive experiments demonstrate that \sys achieves much better explanation performance than the original base model, which is even comparable to the GPT series on benign and malicious shell commands in both English and Chinese. \sys also achieves superior technique and tactic identification performance. However, \sys can be further improved in four aspects:

\textit{Obfuscation.} We have observed that when presented with obfuscated commands, \sys can notify the analyst within the explanation that the commands have indeed been obfuscated. Moreover, \sys can effectively explain commands with payloads obfuscated using techniques such as \textit{Base64} encoding. This success may be attributed to the emerging ability of existing LLMs to comprehend syntactic transformations~\cite{rao2024tricking}, such as string splicing, common encoding, and simple encryption. However, due to the absence of a benchmark for obfuscated commands, we consider a more comprehensive evaluation of the analysis of obfuscated commands as a focus for our future work.

\textit{Shell session.} A shell session consists of continuous commands executed by potential attackers, which contains more information about the attack. Although \sys has already shown satisfying performance on analyzing a standalone shell command and a compound command that consists of multiple commands, extending to shell sessions that consists of sequential shell commands might be more conducive, \textit{e.g.}, for honey-pot analysis and APT analysis.

\textit{Other logs.} \sys is now designed for assisting shell log auditing. A future direction is to extend the system to analyze other kinds of logs, \textit{e.g.}, network logs, database logs, web server logs, \textit{etc.} Extending \sys to fine-tuning the base model on other log modalities requires extensive labelled data which is expensive. A possible solution is through incremental pre-training on unsupervised data of network logs, database logs, web server logs, \textit{etc.} This is our future direction.

\textit{Base model selection.} In this paper, we choose ChatGLM2-6B as our base model and further fine-tune it into \sys for professional command explanation. We choose ChatGLM2-6B because it is the state-of-the-art within open-source bilingual (Chinese-English) LLMs. But a limitation is that ChatGLM2-6B only has 6 billion parameters, which may not be able to beat the performance of larger models, \textit{e.g.}, GPT-3.5-Turbo and GPT-4. Although we are happy to see that \sys has already achieved satisfying performance compared with our baselines in \S\ref{sec:result}, exploration of larger open-source models as the base model may improve the performance further.

\section*{Acknowledgement}
We sincerely thank all the anonymous reviewers for their valuable comments. This work was supported by China NSFC Grant 61925109 and by Ant Group. The authors from Ant Group were supported by the Leading Innovative and Entrepreneur Team Introduction Program of HangZhou (Grant No. TD2020001). Yanjiao Chen is the corresponding author.

\balance
\bibliographystyle{IEEEtranS}
\bibliography{mybib}

\onecolumn
\appendices

\section{Expression Diversification}\label{apx:expression}
{\centering
\begin{tcolorbox}[title = {Template Set of Different Expressions}, , colback=white, halign={left}, width=.6\linewidth]
{\normalsize
$\bullet$~{Can you clarify \texttt{\underline{<command>}}}

$\bullet$~{Please describe \texttt{\underline{<command>}}}

$\bullet$~{Elaborate on \texttt{\underline{<command>}}}

$\bullet$~{Can you give me more details about \texttt{\underline{<command>}}}

$\bullet$~{Could you shed some light on \texttt{\underline{<command>}}}

$\bullet$~{I would like to understand \texttt{\underline{<command>}}}

$\bullet$~{Can you break down \texttt{\underline{<command>}}}

$\bullet$~{Can you make it clear \texttt{\underline{<command>}}}

$\bullet$~{Can you give a rundown \texttt{\underline{<command>}}}

$\bullet$~{Please provide a detailed explanation \texttt{\underline{<command>}}}

$\bullet$~{I would like a detailed explanation \texttt{\underline{<command>}}}

$\bullet$~{Kindly provide a thorough explanation \texttt{\underline{<command>}}}

$\bullet$~{Can you give a detailed explanation \texttt{\underline{<command>}}}

$\bullet$~{Please explain in detail \texttt{\underline{<command>}}}

$\bullet$~{Can you explain it in detail \texttt{\underline{<command>}}}

$\bullet$~{Could you provide a comprehensive explanation \texttt{\underline{<command>}}}

$\bullet$~{Could you go into a detail about the command \texttt{\underline{<command>}}}
}
\end{tcolorbox}}
\vspace{1cm}

\section{Baselines}\label{apx:baseline}

\begin{table}[h]\centering
\setlength{\abovecaptionskip}{0pt}%
\setlength{\belowcaptionskip}{0pt}%
\caption{The Baselines for Behavior Explainer and Intent Identifier.}\label{tab:generative_models}

\resizebox{.5\linewidth}{!}{
\begin{threeparttable}
\normalsize

\setlength{\tabcolsep}{2.mm}{
\begin{tabular}{@{}lccr@{}}
\toprule
Models & Type & Accessibility & \#Parameters \\ \midrule
GPT-3.5-Turbo & Generative & Closed-Source & 175B \\
GPT-4 & Generative & Closed-Source & 100T \\
ChatGLM2-6B & Generative & Open-Source & 6B \\
\bottomrule
\end{tabular}
}


\end{threeparttable}}
\end{table}

\begin{table}[h]\centering

\setlength{\abovecaptionskip}{0pt}%
\setlength{\belowcaptionskip}{10pt}%
\caption{The Baselines for Doc-Augmented Enhancer.}\label{tab:text2vec_models}

\resizebox{.5\linewidth}{!}{
\begin{threeparttable}
\normalsize

\setlength{\tabcolsep}{1.mm}{
\begin{tabular}{@{}lcccr@{}}
\toprule
Models & Type & Accessibility & \#Param. &  Dimension\tnote{\textdagger} \\ \midrule
Sentence-T5$_\mathrm{large}$ & Text2Vec & Open-Source & 335M & 768 \\
GTR-T5$_\mathrm{XL}$ & Text2Vec & Open-Source & 1,240M & 768 \\
SGPT & Text2Vec & Open-Source & 1,300M & 2,048 \\
E5$_\mathrm{large}$ & Text2Vec & Open-Source & 330M & 1,024 \\
\bottomrule
\end{tabular}
}

\begin{tablenotes}[flushleft]
\item[\textdagger] Embedding Dimension: the models encode arbitrary-length text into a fixed-dimensional vector, \textit{e.g.}, 1,024.
\end{tablenotes}

\end{threeparttable}}
\end{table}


\clearpage
\section{Test Set}\label{apx:testset}
\begin{table*}[h]\centering
\setlength{\abovecaptionskip}{0pt}%
\setlength{\belowcaptionskip}{0pt}%
\caption{Test Sets of the Command Explainer.}\label{tab:testset}

\resizebox{0.9\linewidth}{!}{

\begin{threeparttable}

\normalsize
\setlength{\tabcolsep}{0.8mm}{
\begin{tabular}{@{}cccccrr@{}}
\toprule
\multicolumn{1}{c}{Type}    & \multicolumn{1}{c}{Data Source} & Language & Task\tnote{\textdaggerdbl}             & Prompt\tnote{\textdagger}      & \#Samples & Alias       \\ \midrule
\multirow{24}{*}{\makecell[c]{Malicious\\Command-Line}} & \multirow{6}{*}{\texttt{reverse-shell}}  &Chinese      & Explanation      & Original Q.     & 785       & \texttt{rs-zh-ex-or} \\
                            &                                 &Chinese      & Explanation      & Augmented Q.     & 743       & \texttt{rs-zh-ex-au} \\
                            &                                 &Chinese      & Explanation      & Augmented Q. w/ Doc  & 773       & \texttt{rs-zh-ex-ad} \\ \cmidrule(l){3-7} 
                            &                                 &English      & Explanation      & Original Q.     & 954       & \texttt{rs-en-ex-or} \\
                            &                                 &English      & Explanation      & Augmented Q.     & 933       & \texttt{rs-en-ex-au} \\
                            &                                 &English      & Explanation      & Augmented Q. w/ Doc  & 919       & \texttt{rs-en-ex-ad} \\ \cmidrule(l){2-7} 
                            & \multirow{9}{*}{\texttt{metta}}          &Chinese      & Explanation      & Original Q.     & 5         & \texttt{me-zh-ex-or} \\
                            &                                 &Chinese      & Explanation      & Augmented Q.     & 5         & \texttt{me-zh-ex-au} \\
                            &                                 &Chinese      & Explanation      & Augmented Q. w/ Doc  & 8         & \texttt{me-zh-ex-ad} \\ \cmidrule(l){3-7} 
                            &                                 &English      & Explanation      & Original Q.     & 9         & \texttt{me-en-ex-or} \\
                            &                                 &English      & Explanation      & Augmented Q.     & 7         & \texttt{me-en-ex-au} \\
                            &                                 &English      & Explanation      & Augmented Q. w/ Doc  & 10        & \texttt{me-en-ex-ad} \\ \cmidrule(l){3-7} 
                            &                                 &English      & Behavior         & Original Q.     & 196       & \texttt{me-en-be-or} \\
                            &                                 &English      & Behavior         & Augmented Q.     & 212       & \texttt{me-en-be-au} \\
                            &                                 &English      & Behavior         & Augmented Q. w/ Doc  & 323       & \texttt{me-en-be-ad} \\ \cmidrule(l){2-7} 
                            & \multirow{9}{*}{\texttt{atomic-red-team}}       &Chinese      & Explanation      & Original Q.     & 68        & \texttt{rt-zh-ex-or} \\
                            &                                 &Chinese      & Explanation      & Augmented Q.     & 58        & \texttt{rt-zh-ex-au} \\
                            &                                 &Chinese      & Explanation      & Augmented Q. w/ Doc  & 71        & \texttt{rt-zh-ex-ad} \\ \cmidrule(l){3-7} 
                            &                                 &English      & Explanation      & Original Q.     & 66        & \texttt{rt-en-ex-or} \\
                            &                                 &English      & Explanation      & Augmented Q.     & 57        & \texttt{rt-en-ex-au} \\
                            &                                 &English      & Explanation      & Augmented Q. w/ Doc  & 69        & \texttt{rt-en-ex-ad} \\ \cmidrule(l){3-7} 
                            &                                 &English      & Behavior         & Original Q.     & 1464      & \texttt{rt-en-be-or} \\
                            &                                 &English      & Behavior         & Augmented Q.     & 1418      & \texttt{rt-en-be-au} \\
                            &                                 &English      & Behavior         & Augmented Q. w/ Doc  & 1918      & \texttt{rt-en-be-ad} \\ \midrule
\multirow{10}{*}{\makecell[c]{Benign\\Command-Line}}    & \multirow{10}{*}{\texttt{NL2Bash}}       &Chinese      & Explanation      & Original Q.     & 397       & \texttt{nb-zh-ex-or} \\
                            &                                 &Chinese      & Explanation      & Augmented Q.     & 396       & \texttt{nb-zh-ex-au} \\
                            &                                 &Chinese      & Explanation      & Augmented Q. w/ Doc  & 369       & \texttt{nb-zh-ex-ad} \\ \cmidrule(l){3-7} 
                            &                                 &Chinese      & Explanation w/ Doc & Original Q.     & 410       & \texttt{nb-zh-ed-or} \\
                            &                                 &Chinese      & Explanation w/ Doc & Augmented Q.     & 409       & \texttt{nb-zh-ed-au} \\
                            &                                 &Chinese      & Explanation w/ Doc & Augmented Q. w/ Doc  & 411       & \texttt{nb-zh-ed-ad} \\ \cmidrule(l){3-7} 
                            &                                 &Chinese      & Multi-round      & Original Q.     & 5162      & \texttt{nb-zh-mu-or} \\ \cmidrule(l){3-7} 
                            &                                 &English      & Explanation w/ Doc & Original Q.     & 386       & \texttt{nb-zh-ed-or} \\
                            &                                 &English      & Explanation w/ Doc & Augmented Q.     & 396       & \texttt{nb-zh-ed-au} \\
                            &                                 &English      & Explanation w/ Doc & Augmented Q. w/ Doc  & 407       & \texttt{nb-zh-ed-ad} \\\midrule
\makecell[c]{Benign \&\\Malicious}   & \makecell[c]{\texttt{atomic-red-team} \&\\\texttt{NL2Bash}} &\makecell[c]{Chinese \&\\English}      & \makecell[c]{Explanation \&\\Explanation w/ Doc} & Original  & 200       & \texttt{HumanCheck} \\ \bottomrule

\end{tabular}
}

\begin{tablenotes}[flushleft]
\item[\textdaggerdbl] Explanation: detailed explanation; Explanation w/ Doc: detailed explanation given documentations; Multi-round: multi-round interation; Behavior: behavior summarization.
\item[\textdagger] Original Q.: use the same prompts as when requesting ChatGPT; Augmented Q.: use augmented prompts; Augmented Q. w/ Doc: use augmented prompts with documentations.
\end{tablenotes}

\end{threeparttable}}
\end{table*}

\clearpage
\section{Different Expressions of Query}\label{apx:diversified}
\begin{table*}[h]\centering
\setlength{\abovecaptionskip}{0pt}%
\setlength{\belowcaptionskip}{0pt}%
\caption{The Performance on Diversified Queries.}\label{tab:diversified}

\resizebox{\linewidth}{!}{
\begin{threeparttable}

\footnotesize
\setlength{\tabcolsep}{0.8mm}{
\begin{tabular}{@{}l|cccc|cccc|cccc@{}}
\toprule
\multirow{2}{*}[-3pt]{Model} & \multicolumn{4}{c|}{Original Query}                                                                                        & \multicolumn{4}{c|}{Diversified Query\tnote{\textdagger}}                                                                                           & \multicolumn{4}{c}{Diversified Query w/ Doc.\tnote{\textdagger}}                                                                                       \\ \cmidrule(l){2-13} 
                       & \multicolumn{1}{r}{ROUGE-1} & \multicolumn{1}{r}{ROUGE-2} & \multicolumn{1}{r}{ROUGE-$\ell$} & \multicolumn{1}{r|}{BLEU-4} & \multicolumn{1}{r}{ROUGE-1} & \multicolumn{1}{r}{ROUGE-2} & \multicolumn{1}{r}{ROUGE-$\ell$} & \multicolumn{1}{r|}{BLEU-4} & \multicolumn{1}{r}{ROUGE-1} & \multicolumn{1}{r}{ROUGE-2} & \multicolumn{1}{r}{ROUGE-$\ell$} & \multicolumn{1}{r}{BLEU-4} \\ \midrule
GPT-3.5-Turbo          &61.2          & 37.2          & 46.6          & 50.3          & 45.9          & 19.7          & 29.8          & 30.0          & 43.6          & 18.2          & 27.6          & 24.2          \\
GPT-4                  &50.3          & 24.6          & 34.6          & 41.7          & 45.3          & 20.2          & 30.3          & 38.2          & 45.7          & 20.1          & 30.7          & 37.6          \\
ChatGLM2-6B            &49.3          & 23.6          & 31.1          & 38.0          & 41.4          & 17.3          & 26.5          & 31.8          & 42.9          & 17.8          & 26.9          & 32.9          \\
\sys                   &\textbf{\small70.4} & \textbf{\small52.0} & \textbf{\small58.7} & \textbf{\small57.9} & \textbf{\small68.5} & \textbf{\small49.3} & \textbf{\small56.6} & \textbf{\small56.2} & \textbf{\small68.5} & \textbf{\small49.6} & \textbf{\small57.0} & \textbf{\small56.8} \\ \bottomrule
\end{tabular}
}

\begin{tablenotes}[flushleft]
\item[\textdagger] Diversified Query: the form of the query is converted using Equation~\eqref{equ:aug}.
\end{tablenotes}

\end{threeparttable}}
\end{table*}
\raggedbottom

\section{Different Languages}\label{apx:language}
\begin{table}[h]\centering
\setlength{\abovecaptionskip}{0pt}%
\setlength{\belowcaptionskip}{0pt}%
\caption{The Performance on Chinese and English.}\label{tab:language}

\resizebox{.5\linewidth}{!}{
\begin{threeparttable}

\footnotesize
\setlength{\tabcolsep}{2mm}{
\begin{tabular}{@{}l|cccr@{}}
\toprule
                        & \multicolumn{4}{c}{Chinese}                                           \\ 
\multirow{-2}{*}{Model} & ROUGE-1                                  & ROUGE-2 & ROUGE-$\ell$ & BLEU-4 \\ \midrule
GPT-3.5-Turbo           & 46.6                                     & 20.6    & 29.2    & 19.6   \\
GPT-4                   & 46.5                                     & 20.9    & 31.9    & 26.4   \\
ChatGLM2-6B             & 43.5                                     & 18.2    & 26.8    & 21.7   \\
\sys     & \textbf{\small65.7}                                     & \textbf{\small44.0}    & \textbf{\small51.6}    & \textbf{\small42.9}   \\ \midrule\midrule
                        & \multicolumn{4}{c}{English}                                           \\ 
\multirow{-2}{*}{Model} & ROUGE-1                                  & ROUGE-2 & ROUGE-$\ell$ & BLEU-4 \\ \midrule
GPT-3.5-Turbo           & 50.8                                     & 26.9    & 32.5    & 38.6   \\
GPT-4                   & 49.4                                     & 26.1    & 32.5    & 49.1   \\
ChatGLM2-6B             & 48.1                                     & 25.5    & 30.4    & 42.8   \\
\sys     & \textbf{\small70.9}                                     & \textbf{\small53.2}    & \textbf{\small58.4}    & \textbf{\small65.1}   \\ \bottomrule
\end{tabular}
}


\end{threeparttable}}
\end{table}

\end{document}